\documentclass[preprint,review,12pt]{elsarticle}

\usepackage{amssymb}
\usepackage{subcaption}
\usepackage{graphicx}
\usepackage{verbatim}
\usepackage{mathtools}
\usepackage{hyperref}
\usepackage{mathrsfs}
\usepackage{enumitem}
\usepackage{color}
\usepackage{booktabs}
\usepackage{natbib}
\usepackage[algoruled,boxed,lined]{algorithm2e}
\usepackage{algorithmic}
\usepackage[normalem]{ulem}

\graphicspath{{figures/}}
\newcommand{\linesolid}[1][5ex]{\rule[0.5ex]{5ex}{0.2ex}} 
\newcommand{\linedotdash}[1][5ex]{\rule[0.5ex]{2ex}{0.2ex}$\,$\rule[0.5ex]{0.4ex}{0.2ex}$\,$\rule[0.5ex]{2ex}{0.2ex}}
\newcommand{\linedash}[1][5ex]{\rule[0.5ex]{2.2ex}{0.2ex}$\;$\rule[0.5ex]{2.2ex}{0.2ex}}
\newcommand{\linedot}[1][5ex]{\rule[0.5ex]{0.5ex}{0.2ex}$\,$\rule[0.5ex]{0.5ex}{0.2ex}$\,$\rule[0.5ex]{0.5ex}{0.2ex}$\,$\rule[0.5ex]{0.5ex}{0.2ex}$\,$\rule[0.5ex]{0.5ex}{0.2ex}$\,$\rule[0.5ex]{0.5ex}{0.2ex}}

\newcommand{\trans}{{\scriptscriptstyle\mathsf{T}}}

\journal{Journal of Computational Physics}

\begin{document}
	
\begin{frontmatter}	
	
	\title{A realizable second-order advection method with variable flux limiters for moment transport equations}
	
	\author{Byeongyeob Choi }
	\author{Jehyun Baek \corref{cor1}}
	\ead{jhbaek@postech.ac.kr}
	\author{Donghyun You \corref{cor1}}
	\ead{dhyou@postech.ac.kr}
	\address{Department of Mechanical Engineering, Pohang University of Science and Technology, 77 Cheongam-Ro, Nam-Gu, Pohang, Gyeongbuk 37673, Republic of Korea \vspace{-0.4in}}
		
	\cortext[cor1]{Corresponding authors.}
			 
	\begin{abstract}
		A second-order total variation diminishing (TVD) method with variable flux limiters is proposed to overcome the non-realizability issue, which has been one of major obstacles in applying the conventional second-order TVD schemes to the moment transport equations. In the present method, a realizable moment set at a cell face is reconstructed by allowing the flexible selection of the flux limiter values within the second-order TVD region. Necessary conditions for the variable flux limiter scheme to simultaneously satisfy the realizability and the second-order TVD property for the third-order moment set are proposed. The strategy for satisfying the second-order TVD property is conditionally extended to the fourth- and fifth-order moments. The proposed method is verified and compared with other high-order realizable schemes in one- and two-dimensional configurations, and is found to preserve the realizability of moments while satisfying the high-order TVD property for the third-order moment set and conditionally for the fourth- and fifth-order moments.
	\end{abstract}
	
\end{frontmatter}
	
\section{Introduction}  

Multiphase flow systems considering non-inertial elements in a carrier fluid are applied in widespread applications such as soot in combustion, droplets in steam turbines, and aerosol transmission. The number of elements dealt with in these applications is typically so large that tracking individual elements is computationally impractical, requiring an Eulerian approach. In an Eulerian approach, the population of non-inertial elements is generally described with a number density function (NDF) depending on time, spatial coordinates, and one internal coordinate corresponding to the size of elements. The evolution of an NDF can be obtained by solving a population balance equation (PBE) usually coupled with a computational fluid dynamics (CFD) simulation.

When solving a PBE numerically, applying direct discretization over the internal coordinate is undesirable due to the high computational cost \citep{marchisio2013}. Instead, quadrature-based moment methods (QBMMs) \citep{mcgraw1997, marchisio2005, yuan2012} have gained huge popularity as a practical solution method for a PBE because they have proven to be accurate and numerically efficient when combined with CFD simulations \citep{fox2018}. In QBMMs, the NDF is approximated by the sum of weighted kernel density functions centered on the quadrature abscissas. This quadrature approximation allows the NDF to be reconstructed from a finite number of moments. Consequently, the solution of a PBE can be computed by solving a system of moment transport equations, which are similar to the scalar transport equations in CFD simulations. 

However, unlike independent scalars in the scalar transport equations, the moments of an NDF are implicitly correlated to each other because they share the underlying NDF. The set of moments derived from the single NDF should satisfy the mathematical constraint that the determinants of a series of the matrices, which take the form of the Hankel matrix with the moments as the entries, do not become negative. In other words, the set of moments that do not satisfy the mentioned mathematical constraint becomes a set of worthless values which cannot correspond to any NDF. This mathematical constraint is known as the realizability condition, and the set of moments satisfying the realizability condition is considered realizable. In order to derive physically meaningful results from the system of moment transport equations, the realizability condition should be satisfied during the numerical simulation, which makes dealing with moments more complicated than dealing with independent scalars.

The situation where the realizability of moments is problematic occurs mainly in treating the advection terms in the moment transport equations. The realizability condition that the determinants should not be negative limits the ranges of values that each moment can have, and the ranges are determined by other moments constituting the moment set. However, the interpolation processes applied to each moment to reconstruct the advected moments are generally independent of each other and do not consider the realizability condition, resulting in a non-realizable moment set. Therefore, conventional high-order advection schemes are known to have the problem of forming a non-realizable moment set while solving the moment transport equations \citep{wright2007}. Only the first-order scheme guarantees the realizability of the transported moment set \citep{desjardins2008}, but it undergoes excessive numerical diffusion. Therefore, developing a realizable high-order advection scheme for moment transport equations is of particular importance.

In the previous studies, instead of directly reconstructing the moments, the moments at the cell face are indirectly computed from the reconstructed values of other variables related to the moments, such as quadrature weights and abscissas \citep{vikas2011}, canonical moments \citep{Kah2012}, and $\zeta$ variables \citep{laurent2017, passalacqua2020}. The introduced variables can replace the role of the determinants that appeared in the realizability condition, so if these variables are non-negative, the realizability condition is satisfied. This characteristic allows each variable to be reconstructed independently. However, whether the indirectly reconstructed moments satisfy the total variation diminishing (TVD) property is not guaranteed. Accordingly, the boundedness of moments is not guaranteed. Recently, Shiea et al. \citep{shiea2020} overcame the realizability issue in conventional TVD schemes by introducing the equal limiter scheme that equalizes all flux limiters into a minimum limiter value. Therefore, this scheme can guarantee the boundedness property. However, the lowered flux-limiter values may decrease the order of accuracy.

In the present study, a novel second-order TVD method with variable flux limiters is proposed to overcome the non-realizability issue. The key idea is to allow the flexible selection of the flux limiter values within the second-order TVD region to find a combination of moments that satisfies the realizability condition. Therefore, the boundedness of moments can be considered while preserving second-order accuracy and realizability. The proposed method is verified for the transport of the fifth-order moment set, which is the optimal number of moments considering accuracy against efficiency \citep{marchisio2003}, and applicable in practical moment transport problems \citep{gerber2007, vlieghe2016, starzmann2018, ding2021}. The verifications are conducted on one- and two-dimensional configurations compared with those of other realizable high-order schemes, the $\zeta$ simplified scheme \citep{laurent2017} and the equal flux limiter scheme \citep{shiea2020}.

\section{Moment transport equation and realizability} \label{Ch2}

\subsection{Moment transport equation} \label{Ch2-1}
The NDF is denoted as $f(t,\mathbf{x},\xi)$, where $t$ is time, $\mathbf{x}=\left(x_1, x_2, x_3\right)$ is spatial coordinate, and $\xi$ is an internal coordinate. In the present study, the internal coordinate $\xi$ represents the size of dispersed phase elements (e.g., particles and droplets) on the interval $[0,+\infty)$. In the case of pure advection in the carrier fluid, the PBE is written as follows:
\begin{equation}
	\frac{\partial{f}}{\partial{t}}+div\left(\mathbf{u}f\right)=0,
	\label{Eq_1}
\end{equation}
where $\mathbf{u}=\left(u_1, u_2, u_3\right)^\trans$ is the velocity vector of the carrier fluid, and the superscript $\trans$ denotes matrix transpose.

In moment methods, the NDF is approximated as a finite set of moments. Then, the PBE can be reformulated into a system of moment transport equations. The $k$th-order moment $m_k$ is defined as 
\begin{equation}
	m_{k}(t, \mathbf{x})=\int_{0}^{+\infty} \xi^{k} f(t, \mathbf{x}, \xi) \mathrm{d} \xi.
	\label{Eq_2}
\end{equation}
Subsequently, the system of moment transport equations is  written as follows:
\begin{equation}
	\frac{\partial{\mathbf{m}_{N}}}{\partial{t}}+div\left(\mathbf{m}_{N}\otimes\mathbf{u}\right)=0,
	\label{Eq_3}
\end{equation}
where $\mathbf{m}_N =\left(m_0,\ldots, m_N\right)^\trans$ is the $N$th-order moment set.

The moment transport equations are decoupled from each other. However, a physically valid range of the $k$th-order moment is restricted by the lower-order moments up to the $(k-1)$th-order to satisfy the realizability condition. Therefore, the independently transported moments without considering the realizability condition are possibly corrupted and represents an unphysical NDF. This corrupted moment set causes the non-realizability issue, leading to the crucial problem of failure in moment-inversion algorithms \citep{gordon1968, wheeler1974} to close the moment equations. Hence, the non-realizability problem is the major obstacle to practical applications of moment methods.

\subsection{Realizability condition} \label{Ch2-2}
The realizability condition is based on physical constraints of the NDF. For example, the zeroth-order moment is not allowed to be negative due to the condition of non-negative number density. Furthermore, the variance (i.e.,  $\sigma^{2}=m_2-m_1^2/m_0$) of the NDF must be non-negative. Therefore, it imposes another constraint on constructing a moment set, $m_2m_0-m_1^2\ge0$. In the theory of moments, it is proved that the realizability condition can be defined using the Hankel determinants \citep{dette1997, shohat1943, gautschi2004}. 

The Hankel determinants are defined as follows:
\begin{equation}
	\Delta_{p+2l}=\left|\begin{array}{cccc}
	m_{p} & m_{p+1} &\cdots & m_{p+l} \\
	m_{p+1} & m_{p+2} & \cdots & m_{p+l+1} \\
	\vdots & \vdots & \vdots & \vdots \\
	m_{p+l} & m_{p+l+1} &\cdots & m_{p+2l}
	\end{array}\right|,
	\label{Eq_4}
\end{equation}
for $p=0$ and $1$, and $l\ge 0$. The moment set is included in the interior of the moment space if the Hankel determinants are all positive when the NDF is supported on $[0,+\infty)$. To be specific, the $N$th-order moment set $\mathbf{m}_N =\left(m_0,\ldots, m_N\right)^\trans$ is strictly realizable if $\Delta_{k}>0$ for $k=0,1,\ldots,N$. A part of the Hankel determinants becomes 0 if the moment set belongs to the boundary of the moment space. Note that $\Delta_n=0$ implies that $\Delta_k=0$ for all $k\ge{n+1}$. Therefore, if there exists $n\le N$, the Hankel determinants of $\mathbf{m}_N$ are given as follows:
\begin{equation}
	\Delta_{0}>0,\ldots,\Delta_{n-1}>0,\Delta_{n}=0,\ldots,\Delta_{N}=0.
	\label{Eq_5}
\end{equation}
The smallest $n$ in which the Hankel determinant becomes 0 is denoted by $\mathcal{N}(\mathbf{m}_{N})$. Then, the realizability condition of $\mathbf{m}_N$ can be readily represented in computational algorithms with $\mathcal{N}(\mathbf{m}_{N})>N$ for the strictly realizable case and $\mathcal{N}(\mathbf{m}_{N})\le N$ for the boundary belonging case.

\subsection{Finite volume method} \label{Ch2-3}
In this section, the realizability issues in the finite volume methods (FVMs) are discussed. In the context of FVMs, the integral form of the $k$th-order moment transport equation is written as follows:
\begin{equation}
\frac{\mathrm{d}{m_{k,i}}}{\mathrm{d}{t}}+\frac{1}{\Delta V_{i}}\sum_{e}\left[\left(\mathbf{u}_{e}\cdot{\mathbf{S}}_{e}\right)m_{k,e}\right]=0,
\label{Eq_6}
\end{equation}
where the subscripts $i$ and $e$ are the indices of computational cells and their faces, respectively. Therefore, $\Delta V_{i}$ refers to the volume of cell $i$, and $\mathbf{u}_{e}$ and $\mathbf{S}_{e}$ refer to the velocity vector and the surface vector at face $e$, respectively. The second term in Eq. \eqref{Eq_6} is the net flux of the moment (i.e., inflow and outflow through the cell faces). In FVMs, the moment at the face is not available directly and is generally interpolated from the values stored in neighboring cells. 

Let us consider the conventional second-order TVD schemes. For the sake of simplicity, a uniformly distributed one-dimensional grid and positive velocity (i.e., $u>0$) are assumed. Then, the moments of upwind and downwind neighboring cells of face $e$ are denoted as $m_{k,i}$ and $m_{k,i+1}$, respectively. The moment reconstructed at the face is expressed as follows:
\begin{equation}
	\begin{gathered}
	m_{k,e}=m_{k,i} + \lambda_{e}\phi\left(r_{k,e}\right)\left(m_{k,i+1}-m_{k,i}\right), \\
	r_{k,e}=\frac{m_{k,i}-m_{k,i-1}}{m_{k,i+1}-m_{k,i}}, \\
	 \lambda_{e}=\frac{1}{2},
	\label{Eq_7}
	\end{gathered}
\end{equation}
where $\phi(r)$ is the flux limiter function. The flux limiter can control the contribution of the anti-diffusive term $(m_{k,i+1}-m_{k,i})$. The TVD property is used when choosing the flux limiter to prevent oscillation. Fig. \ref{Fig_1} shows Sweby's TVD diagram \citep{sweby1984}, graphically showing the region satisfying the TVD property. The shaded area refers to the TVD region, and the hatched area refers to the second-order region. The simplest choice of the flux limiter function in the second-order TVD region can be the minmod flux limiter \citep{roe1986}:
\begin{equation}
	\phi_{mm}(r)=\max\left[0,\min\left(1,r\right)\right].
	\label{Eq_8}
\end{equation}

The moment set $\mathbf{m}_{N,e}=\left(m_{0,e},\ldots,m_{1,e}\right)^\trans$ reconstructed at the face as follows by rearranging Eq. \eqref{Eq_7}:
\begin{equation}
	\left[\begin{array}{c}
	m_{0,e} \\
	m_{1,e} \\
	\vdots \\
	m_{N,e}
	\end{array}\right]
	=\left[\begin{array}{c}
	{\left(1-\lambda_{e} \phi\left(r_{0,e}\right)\right) m_{0,i}} \\
	{\left(1-\lambda_{e} \phi\left(r_{1,e}\right)\right) m_{1,i}} \\
	\vdots \\
	{\left(1-\lambda_{e} \phi\left(r_{N,e}\right)\right) m_{N,i}}
	\end{array}\right]	    
	+\left[\begin{array}{c}
	{\lambda_{e} \phi\left(r_{0,e}\right)m_{0,i+1}} \\
	{\lambda_{e} \phi\left(r_{1,e}\right)m_{1,i+1}} \\
	\vdots \\
	{\lambda_{e} \phi\left(r_{N,e}\right)m_{N,i+1}}
	\end{array}\right].
	\label{Eq_9}	 
\end{equation}
Even if two moment sets $\mathbf{m}_{N,i}$ and $\mathbf{m}_{N,i+1}$ are realizable, the reconstructed moment set $\mathbf{m}_{N,e}$ can be non-realizable because the value of the flux limiters $\phi\left(r_{k,e}\right)$ can vary for each order of the moment. Shiea et al. \citep{shiea2020} introduced the concept of the equal flux limiter to overcome the non-realizability problem, where all the flux limiter values are equalized into their minimum value as follows:
\begin{equation}
	\phi_{e}^{\min} = \min\left(\phi\left(r_{0,e}\right),\ldots,\phi\left(r_{N,e}\right) \right).
	\label{Eq_10}
\end{equation}
Accordingly, Eq. \eqref{Eq_9} can be rewritten by substituting the minimum value $\phi_{e}^{\min}$ into the flux limiters $\phi\left(r_{k,e}\right)$:
\begin{equation}
	\left[\begin{array}{c}
	m_{0,e} \\
	m_{1,e} \\
	\vdots \\
	m_{N,e}
	\end{array}\right]
	=\left(1-\lambda_{e} \phi_e^{\min}\right) \left[\begin{array}{c}
	{m_{0,i}} \\
	{m_{1,i}} \\
	\vdots \\
	{m_{N,i}}
	\end{array}\right]	    
	+\lambda_{e} \phi_e^{\min}\left[\begin{array}{c}
	{m_{0,i+1}} \\
	{m_{1,i+1}} \\
	\vdots \\
	{m_{N,i+1}}
	\end{array}\right].	
	\label{Eq_11} 
\end{equation}
Then, the moment set at the face $\mathbf{m}_{N,e}$ becomes the convex combination of the two moment sets $\mathbf{m}_{N,i}$ and $\mathbf{m}_{N,i+1}$ because $\phi_{e}^{\min}$ is on the interval $[0,2]$. The moment space is closed under addition. Consequently, their convex combination $\mathbf{m}_{N,e}$ is guaranteed to be realizable if $\mathbf{m}_{N,i}$ and $\mathbf{m}_{N,i+1}$ are all realizable.

The $\zeta$ simplified scheme \citep{laurent2017} is another realizable scheme. In this scheme, a variable $\zeta$ corresponding to the moment space is employed to overcome the difficulty in the use of the Hankel determinants directly for the moment reconstruction at the face. The $k$-th order $\zeta$ value $\zeta_k$ can be defined using the Hankel determinants as follows \citep{dette1997}:
\begin{equation}
	\zeta_{k}=\frac{\Delta_{k} \Delta_{k-3}}{\Delta_{k-1}\Delta_{k-2}},
	\label{Eq_12}
\end{equation}
where $\Delta_{k}:=1$ if $k\le0$. Then, $\zeta_k$ can be interpreted as the other  form of the Hankel determinant $\Delta_{k}$. Consequently, the realizability condition of the moment set $\mathbf{m}_{N}$ in Section \ref{Ch2-2} can be rewritten with the $\zeta$ set $\mathbf{Z}_{N}=(\zeta_0,...,\zeta_N)^\trans$. For $k=0,1,...,N$, the moment set $\mathbf{m}_{N}$ is strictly realizable if $\zeta_k>0$. If it belongs to the boundary of the moment space and there exists $n\le N$, then:
\begin{equation}
	\zeta_{0}>0,\ldots,\zeta_{n-1}>0,\;\zeta_{n}=0,\ldots,\zeta_{N}=0.
	\label{Eq_13}
\end{equation}
The moment $m_k$ can be written as a function of $\zeta_k$ as follows:
\begin{equation}
	m_k=\zeta_0\left[P_k\left(\zeta_1,\ldots,\zeta_{k-1}\right)+\prod_{j=1}^{k}{\zeta_j}\right],
	\label{Eq_14}
\end{equation}
where
    {
	
	\centering{$P_1=0,$}
	
	\centering{$P_2\left(\zeta_1\right) = \zeta_1^2,$}
	
	\centering{$P_k\left(\zeta_1,\ldots,\zeta_{k-1}\right) = \zeta_1\left[\left(\zeta_1+\zeta_2\right)^{k-1}+\zeta_2\zeta_3 Q_k\left(\zeta_1,\ldots,\zeta_{k-1}\right)\right] \; \textrm{for} \; k\ge3,$}
	
	} 
\noindent with
	{	
	
	\centering{$Q_3\left(\zeta_1,\zeta_2\right) = 0,$}
	
	\centering{$Q_4\left(\zeta_1,\zeta_2,\zeta_{3}\right) = 2\left(\zeta_1+\zeta_2\right)+\zeta_3,$}
	
	\centering{$Q_5\left(\zeta_1,\ldots,\zeta_{4}\right) = 3\left(\zeta_1+\zeta_2\right)^2+2\left(\zeta_1+\zeta_2\right)\left(\zeta_3+\zeta_4\right)+\left(\zeta_3+\zeta_4\right)^2+\zeta_2\zeta_3.$}
	
	}	
\noindent The detailed process is referred to in \citep{laurent2017}. Then, it is possible to reconstruct the moments indirectly by reconstructing $\zeta_{k,e}$ at the face. The reconstructed $\zeta_{k,e}$ at the face is given by
\begin{equation}
	\begin{gathered}
	\zeta_{k,e}=\zeta_{k,i} + \lambda_{e}\phi\left(r_{k,e}^{\zeta}\right)\left(\zeta_{k,i+1}-\zeta_{k,i}\right), \\
	r_{k,e}^{\zeta}=\frac{\zeta_{k,i}-\zeta_{k,i-1}}{\zeta_{k,i+1}-\zeta_{k,i}}, \\
	\lambda_{e}=\frac{1}{2},
	\end{gathered}
	\label{Eq_15}
\end{equation}
where $\phi(r)$ is the flux limiter function. $\zeta$ set $\mathbf{Z_{N,e}}=(\zeta_{0,e},\ldots,\zeta_{N,e})$ can be written in the similar form of Eq. \eqref{Eq_9} as follows:
\begin{equation}
	\left[\begin{array}{c}
	\zeta_{0,e} \\
	\zeta_{1,e} \\
	\vdots \\
	\zeta_{N,e}
	\end{array}\right]
	=\left[\begin{array}{c}
	{\left(1-\lambda_{e} \phi\left(r_{0,e}^{\zeta}\right)\right) \zeta_{0,i}} \\
	{\left(1-\lambda_{e} \phi\left(r_{0,e}^{\zeta}\right)\right) \zeta_{1,i}} \\
	\vdots \\
	{\left(1-\lambda_{e} \phi\left(r_{0,e}^{\zeta}\right)\right) \zeta_{N,i}}
	\end{array}\right]	    
	+\left[\begin{array}{c}
	{\lambda_{e} \phi\left(r_{0,e}^{\zeta}\right)\zeta_{0,i+1}} \\
	{\lambda_{e} \phi\left(r_{0,e}^{\zeta}\right)\zeta_{1,i+1}} \\
	\vdots \\
	{\lambda_{e} \phi\left(r_{0,e}^{\zeta}\right)\zeta_{N,i+1}}
	\end{array}\right].	 
	\label{Eq_16}
\end{equation}
The realizability is determined by the non-negativity of $\zeta_{k,e}$. If the moment sets $\mathbf{m}_{N,i}$ and $\mathbf{m}_{N,i+1}$ are realizable, the corresponding $\zeta$ sets $\mathbf{Z}_{N,i}$ and $\mathbf{Z}_{N,i+1}$ have all non-negative components. Therefore, the components of $\mathbf{Z}_{N,e}$ are guaranteed to be non-negative because the coefficients $\left(1-\lambda_{e} \phi\left(r_{k,e}^{\zeta}\right)\right)$ and $\lambda_{e} \phi\left(r_{k,e}^{\zeta}\right)$ are also non-negative due to their range. When the moment set belongs to the boundary of the moment space, $\zeta_{n,e}=0$ for the case of $n\le N$ leads to that of $\zeta_{k,e}=0$ for all $k\ge{n+1}$. Finally, the realizable moment set is obtainable at the face by converting the reconstructed $\zeta$ set.

Here, we return to the fully described moment transport equation. Until now, we discussed the realizability of only the reconstructed moment set at the cell face. However, this requirement alone is not sufficient to guarantee the realizability of the overall moment transport process. For a detailed discussion, the fully discretized form of the system of the moment transport equations is derived by applying the explicit Euler scheme to Eq. \eqref{Eq_6} as follows: 
\begin{equation}
	\mathbf{m}_{N,i}^{n+1}=\mathbf{m}_{N,i}^{n}-\frac{\Delta t}{\Delta V_{i}}\sum_{e}\left[\left(\mathbf{u}_{e}\cdot{\mathbf{S}}_{e}\right)\mathbf{m}_{N,e}^{n}\right].
	\label{Eq_17}
\end{equation}
In Eq. \eqref{Eq_17}, the second term on the right-hand side can be separated into two parts of inflow and outflow faces, which is expressed as 
\begin{equation}
	\mathbf{m}_{N,i}^{n+1}=\mathbf{m}_{N,i}^{n}
	+\underbrace{\frac{\Delta t}{\Delta V_{i}}\sum_{ei=1}^{N_{ei}}\left[\left(-\mathbf{u}_{ei}\cdot{\mathbf{S}}_{ei}\right)\mathbf{m}_{N,ei}^{n}\right]}_{\textrm{inflow}}
	-\underbrace{\frac{\Delta t}{\Delta V_{i}}\sum_{eo=1}^{N_{eo}}\left[ \left(\mathbf{u}_{eo}\cdot{\mathbf{S}}_{eo}\right)\mathbf{m}_{N,eo}^{n}\right]}_{\textrm{outflow}},
	\label{Eq_18}
\end{equation}
where the subscripts $ei$ and $eo$ denote the indices of inflow and outflow faces, respectively. In addition, $N_{ei}$ and $N_{eo}$ denote the numbers of inflow and outflow faces, respectively. 

As mentioned previously, the moment space is closed under addition. Therefore, the inflow term does not cause the non-realizability problem. However, the moment space is not closed under subtraction, so the outflow term can cause the non-realizability problem. The time step size $\Delta{t}$ in Eq. \eqref{Eq_18} cannot be determined explicitly because the non-realizability problem can occur even at a tiny time step size. Therefore, Laurent and Nguyen \citep{laurent2017} suggested the method to explicitly obtain a time step size by splitting the computational cell into three parts from the idea of Berthon \citep{berthon2005}, and it is extended to the cases of unstructured grids by Passalacqua et al. \citep{passalacqua2020}. In the method of Passalacqua \citep{passalacqua2020}, the moment set $\mathbf{m}_{N}^{*}$ is newly defined as follows:
\begin{equation}
	\mathbf{m}_{N}^{*}=\left(1+N_{eo}\right)\mathbf{m}_{N,i}^{n}
	-\sum_{eo=1}^{N_{eo}}\mathbf{m}_{N,eo}^{n},
	\label{Eq_19}
\end{equation}
where $N_{eo}$ is the number of outflow faces. Then, the moment set $\mathbf{m}_{N,i}^{n}$ is expressed using $\mathbf{m}_{N}^{*}$ as 
\begin{equation}
	\mathbf{m}_{N,i}^{n}=\frac{1}{1+N_{eo}}\mathbf{m}_{N}^{*}
	+\frac{1}{1+N_{eo}}\sum_{eo=1}^{N_{eo}}\mathbf{m}_{N,eo}^{n}.
	\label{Eq_20}
\end{equation}
Consequently, the following form can be driven by substituting Eq. \eqref{Eq_20} into Eq. \eqref{Eq_18}:
\begin{equation}
	\begin{aligned}
	\mathbf{m}_{N,i}^{n+1}=&\frac{1}{1+N_{eo}}\mathbf{m}_{N}^{*}+\frac{\Delta t}{\Delta V_{i}}\sum_{ei=1}^{N_{ei}}\left[-\left(\mathbf{u}_{ei}\cdot{\mathbf{S}}_{ei}\right)\mathbf{m}_{N,ei}^{n}\right] \\
	 & +\sum_{eo=1}^{N_{eo}}\left[\left(\frac{1}{1+N_{eo}}-\frac{\Delta t}{\Delta V_{i}}\mathbf{u}_{eo}\cdot{\mathbf{S}}_{eo}\right)\mathbf{m}_{N,eo}^{n}\right].
	\end{aligned}
	\label{Eq_21}
\end{equation}
In Eq. \eqref{Eq_21}, $\mathbf{m}_{N,i}^{n+1}$ is ensured to be realizable if the newly defined moment set $\mathbf{m}_{N}^{*}$ is realizable and the coefficients of $\mathbf{m}_{N,eo}^{n}$ are all non-negative. Therefore, the time step can be decided from the following condition that is similar to the Courant-Friedrichs-Lewy (CFL) number. 
\begin{equation}
	\max_{i}\left[\max_{eo}\left(\frac{1}{\Delta V_{i}}\mathbf{u}_{eo}\cdot{\mathbf{S}}_{eo}\right)\right]\Delta t \le \min_{i}\left({\frac{1}{1+N_{eo}}}\right),
	\label{Eq_22}
\end{equation}

The remaining task is to ensure the realizability of $\mathbf{m}_{N}^{*}$. If $\mathbf{m}_{N}^{*}$ is non-realizable, the outflow moment sets $\mathbf{m}_{N,eo}^{n}$ should be modified to make $\mathbf{m}_{N}^{*}$ realizable. Let us consider the $\zeta$ set  $\mathbf{Z}_{N,eo}^{n}$ that corresponds to $\mathbf{m}_{N,eo}^{n}$. The modified value of $\zeta_{k,eo}^{n}$ can be written as follows:
\begin{equation}
	\zeta_{k,eo}^{mod}=\zeta_{k,i}^{n}+D_{k}\left(\zeta_{k,eo}^{n}-\zeta_{k,i}^{n}\right),
	\label{Eq_23}	
\end{equation}
where $D_k$ is the controllable coefficient for the $k$th-order $\zeta$ on the interval $[0,1]$, which is shared in all outflow faces. The $\zeta_{k,eo}^{mod}$ is equivalent to $\zeta_{k,eo}^{n}$ for $D_k=1$, and it becomes identical to $\zeta_{k,i}^{n}$ for $D_k=0$. Therefore, the different moment sets $\mathbf{m}_{N,eo}^{mod}=\left(m_{0,eo}^{mod},\dots,m_{N,eo}^{mod}\right)^\trans$ converted from the $\mathbf{Z}_{N,eo}^{mod}=\left(\zeta_{0,eo}^{mod},\dots,\zeta_{N,eo}^{mod}\right)^\trans$ can be used for $\mathbf{m}_{N}^{*}$ instead of $\mathbf{m}_{N,eo}^{n}$ by reducing the values of $D_k$. Then, the values of $D_k$ are sequentially reduced until the moment set $\mathbf{m}_{N}^{*}$ is realizable under the condition of $\mathcal{N}\left(\mathbf{m}_{N}^{*}\right)\ge \mathcal{N}\left(\mathbf{m}_{N,i}^{n}\right)$. In the worst-case scenario, all values of $D_k$ are reduced to 0. Thus, all $\mathbf{m}_{N,eo}^{mod}$ become identical to $\mathbf{m}_{N,i}^{n}$, and consequently, $\mathbf{m}_{N}^{*}$ becomes $\mathbf{m}_{N,i}^{n}$ ensuring the realizability of $\mathbf{m}_{N}^{*}$. The detailed process is provided in Algorithm \ref{alg_1}.

\section{Variable flux limiter} \label{Ch3}

\begin{figure}
	\centering
	\includegraphics[width=0.5\textwidth]{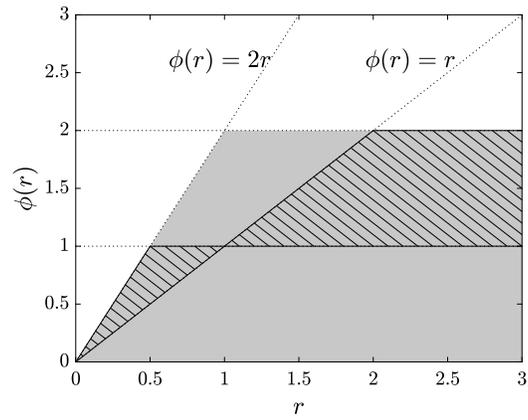}
	\caption{Sweby's TVD diagram for flux limiter functions \citep{sweby1984}. The shaded area refers to the TVD region, and the hatched area refers to the second-order TVD region.}
	\label{Fig_1}
\end{figure}

\begin{figure}
	\centering
	\includegraphics[width=\textwidth]{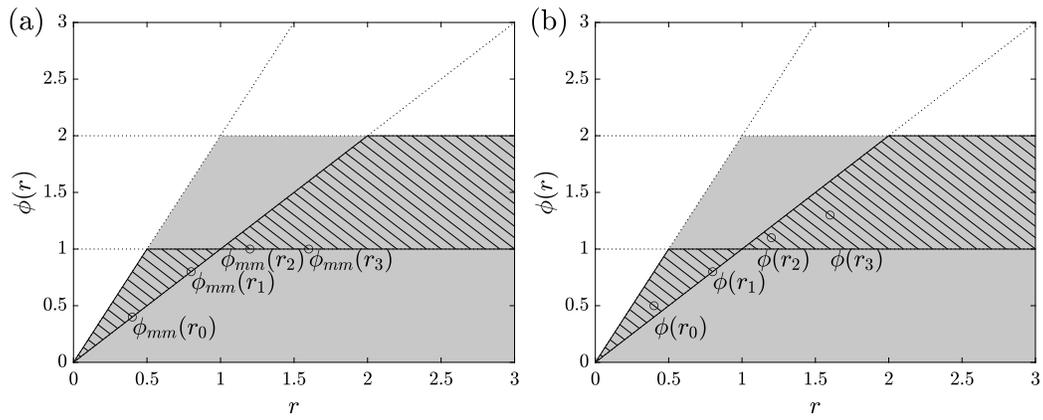}
	\caption{Flux limiter values obtained by applying (a) the minmod flux limiter and (b) the multiple flux limiters.}
	\label{Fig_2}
\end{figure}

As discussed in Section \ref{Ch2-3}, the reconstructed moment set at the face of the computational cell using the conventional second-order TVD schemes can be non-realizable. However, if several flux limiters, not a single limiter, are applied to the reconstruction of the moments, whole the second-order TVD region shown in Fig. \ref{Fig_1} can be utilized, and therefore it may be possible to get a realizable moment set. For example, the values of the minmod flux limiter $\phi_{mm}$ are fixed by their functions, as shown in Fig \ref{Fig_2}(a). Thus, there is no way to avoid the non-realizability problem when it occurs. However, the face moment set can be reconstructed flexibly to satisfy the realizability if each moment can select a different flux limiter as needed. This case is graphically shown in Fig \ref{Fig_2}(b). 

Alternatively, the second-order TVD region in Fig. \ref{Fig_1} can be covered by combining two different flux limiters. These two flux limiters are denoted as $\phi_{lower}$ and $\phi_{upper}$ to represent the lower and upper boundaries. In the present study, the minmod limiter and the superbee limiter are selected as functions of the $\phi_{lower}$ and $\phi_{upper}$, respectively. However, other combinations are also possible to narrow down the covered range. The minmod and superbee limiters are denoted as $\phi_{mm}$ and $\phi_{sb}$, respectively, and are defined as follows \citep{roe1986}: 
\begin{equation}
	\begin{gathered}
	\phi_{lower}(r)=\phi_{mm}(r)=\max\left[0,\min\left(1,r\right)\right], \\
	\phi_{upper}(r)=\phi_{sb}(r)=\max\left[0,\min\left(1,2r\right),\min\left(2,r\right)\right].
	\end{gathered}
	\label{Eq_24}
\end{equation}

In the present study, the two different flux limiters are linearly combined to construct a new variable flux limiter method as follows:
\begin{equation}
	\phi_{var}\left(r,\gamma\right)=\left(1-\gamma\right)\phi_{lower}(r) + \gamma\phi_{upper}(r),\quad 0\le\gamma\le 1.
	\label{Eq_25}
\end{equation}
The reconstructed $k$th-order moment value at the face $e$ using the variable flux limiter can vary based on $\gamma_{k,e}$ as follows: 
\begin{equation}
	\begin{gathered}
	m_{k,e}^{var}\left(\gamma_{k,e}\right)=m_{k,U} + \lambda_e\phi_{var}\left(r_{k,e},\gamma_{k,e}\right)\left(m_{k,D}-m_{k,U}\right), \\
	r_{k,e}=\frac{2\nabla{m_{k,U}\cdot \mathbf{r}_{UD}}}{m_{k,D}-m_{k,U}}-1, \\
	 \lambda_{e}=\frac{1}{2},
	\end{gathered}
	\label{Eq_26}	
\end{equation}
where the subscripts $U$ and $D$ denote the unwind and the downwind neighboring cells of face $e$, respectively. The $\mathbf{r}_{UD}$ is the vector from $U$ to $D$. The maximum and the minimum values of the available $m_{k,e}^{var}$ are denoted as $m_{k,e}^{\max}$ and $m_{k,e}^{\min}$, respectively:
\begin{equation}
	\begin{gathered}
	m_{k,e}^{\max} = \max{\left[m_{k}^{var}\left(\gamma_{k,e}\right)\right]},\\
	m_{k,e}^{\min} = \min{\left[m_{k}^{var}\left(\gamma_{k,e}\right)\right]}.
	\end{gathered}
	\label{Eq_27}
\end{equation}
Accordingly, the value of $m_{k,e}^{var}$ exists in the interval $\left[m_{k,e}^{\min},m_{k,e}^{\max}\right]$. For the sake of simplicity, the superscript $var$ and the subscript $e$ are omitted in the forthcoming discussion. 

This section aims to reconstruct a realizable fifth-order moment set $\mathbf{m}_5 =\left(m_0,\ldots, m_5\right)^\trans$ based on the study that the optimal number of moments considering accuracy against efficiency is six \citep{marchisio2003}. Considering the process of reconstructing $\mathbf{m}_5$ in two stages, $\mathbf{m}_3$ is reconstructed first and then the remaining moments $m_4$ and $m_5$ are obtained.

The realizability condition of the third-order moment set $\mathbf{m}_3 =\left(m_0,\ldots, m_3\right)^\trans$ is provided by two Hankel determinants $\Delta_{2}$ and $\Delta_{3}$: 
\begin{equation}
	\Delta_{2}=\left|\begin{array}{cc}
	m_{0} &  m_{1} \\
	m_{1} &  m_{2}
	\end{array}\right| \ge 0 \quad \textrm{and} \quad
	\Delta_{3}=\left|\begin{array}{cc}
	m_{1} &  m_{2} \\
	m_{2} &  m_{3}
	\end{array}\right| \ge 0.
	\label{Eq_28}
\end{equation}
These conditions can be rewritten as 
\begin{equation}
	\frac{(m_1)^2}{m_0} \le m_2 \quad \textrm{and} \quad \frac{(m_2)^2}{m_1} \le m_3.
	\label{Eq_29}
\end{equation}
Accordingly, the moments that simultaneously satisfy the second-order TVD and the realizability condition are obtained in the following ranges:
\begin{equation}
	\begin{gathered}
	m_0^{\min} \le m_0 \le m_0^{\max}, \\
	m_1^{\min} \le m_1 \le m_1^{\max}, \\
	\max\left[m_2^{\min},\frac{(m_1)^2}{m_0}\right] \le m_2 \le m_2^{\max},  \\
	\max\left[m_3^{\min},\frac{(m_2)^2}{m_1}\right] \le m_3 \le m_3^{\max}. \\
	\end{gathered}
	\label{Eq_30}
\end{equation}
However, the feasible ranges of $m_2$ and $m_3$ depends on the values of the lower-order moments. In other words, possible values of $m_2$ and $m_3$ may not exist depending on the selected values of the lower-order moments. Thus, additional constraints for determining the values of $m_0$, $m_1$, and $m_2$ are introduced to ensure the existence of the higher-order moments. 


Let us consider the range of $m_3$. In order for $m_3$ to exist in the range proposed in Eq. \eqref{Eq_30}, the following condition should be satisfied:
\begin{equation}
	\max\left[m_3^{\min},\frac{(m_2)^2}{m_1}\right]\le m_3^{\max} \quad \Longrightarrow \quad m_2 \le \sqrt{m_1 m_3^{\max}}.
	\label{Eq_31}
\end{equation}
Accordingly, Eq. \eqref{Eq_31} becomes the additional constraint in determining $m_2$ considering the existence of $m_3$. Therefore, the feasible range of $m_2$ is revised as follows:
\begin{equation}
	\max\left[m_2^{\min},\frac{(m_1)^2}{m_0}\right] \le m_2 \le \min\left[m_2^{\max},\sqrt{m_1 m_3^{\max}}\right].
	\label{Eq_32}
\end{equation}
If the same consideration is applied to determine the ranges of $m_1$ and $m_0$, the following ranges can be finally obtained:
\begin{equation}
	\begin{gathered}
		\max \left[{m_0}^{\min}, (1+\varepsilon)\frac{\left(m_{1,*}^{\min}\right)^{2}}{m_{2}^{\max}}, 
		(1+\varepsilon)\sqrt{\frac{\left(m_{1,*}^{\min}\right)^{3}}{m_{3}^{\max}}}\right] \le m_{0} \le m_{0}^{\max}, \\
		m_{1,*}^{\min} \le m_1 
		\le \min\left[m_1^{\max},\sqrt{\frac{m_0 m_2^{\max}}{1+\varepsilon}},\sqrt[3]{\frac{\left(m_0\right)^2 m_3^{\max}}{(1+\varepsilon)^2}}\right], \\
		\max\left[m_2^{\min},(1+\varepsilon)\frac{(m_1)^2}{m_0}\right] \le m_2 \le \min\left[m_2^{\max},\sqrt{m_1 m_3^{\max}}\right], \\
		\max\left[m_3^{\min},\frac{(m_2)^2}{m_1}\right] \le m_3 \le m_3^{\max}, \\
	\end{gathered}
	\label{Eq_33}
\end{equation}
where $m_{1,*}^{\min}=\max\left[m_1^{\min},\left(m_2^{\min}\right)^2/m_3^{\max}\right]$. $\varepsilon$ is introduced to prevent $m_3$ from being forced out of the given range in Eq. \eqref{Eq_33} under the condition of $\Delta_2=0$, and the value of $\varepsilon$ is defined as follows.
\begin{equation}
	\varepsilon = \left\{
	\begin{array}{cc}
		0       & \quad \textrm{if} \quad \sqrt{m_1 m_3^{\min}}\le m_1^2/m_0 \\
		10^{-6} & \quad \textrm{if} \quad\sqrt{m_1 m_3^{\min}} > m_1^2/m_0 \\
	\end{array}.
	\right.
	\label{Eq_34}
\end{equation}
The detailed process for deriving the conditions of Eqs. \eqref{Eq_33} and \eqref{Eq_34} is provided in \ref{Appendix_A}.


In order for the moment set  $\mathbf{m}_3$ to be obtainable under the constraints of Eq. \eqref{Eq_33}, the shaded region in Fig. \ref{Fig_3} should not be empty. Therefore, the following two conditions should be satisfied for the constraints of Eq. \eqref{Eq_33} to be applicable:
\begin{equation}
	m_{1,*}^{\min}\le m_1^{\max} \quad \Longrightarrow \quad \frac{\left(m_2^{\min}\right)^2}{m_3^{\max}}\le m_{1}^{\max},
	\label{Eq_35}	
\end{equation}
\begin{equation}
	\max \left(\left(1+\varepsilon\right)\frac{\left(m_{1,*}^{\min}\right)^{2}}{m_{2}^{\max}}, \left(1+\varepsilon\right)\sqrt{\frac{\left(m_{1,*}^{\min}\right)^{3}}{m_{3}^{\max}}}\right) \le m_{0}^{\max}.
	\label{Eq_36} 
\end{equation}
The values of the moments can be determined sequentially in the order of $m_0$, $m_1$, $m_2$, and $m_3$ under the constraints of Eq. \eqref{Eq_33} if the conditions of Eqs. \eqref{Eq_35} and \eqref{Eq_36} are both satisfied. In addition, the reconstructed moment set $\mathbf{m}_3$ is realizable and definitely preserves the second-order accuracy and the TVD property. The valid range of $m_k$ can be rewritten using the representation of Eq. \eqref{Eq_26} as follows:
\begin{equation}
	\begin{gathered}
	\min\left[m_k\left(\gamma_k^{\min}\right), m_k \left(\gamma_k^{\max}\right)\right] \le m_k\left(\gamma_k\right) \le \max\left[m_k\left(\gamma_k^{\min}\right), m_k\left(\gamma_k^{\max}\right)\right], \\ \gamma_k^{\min} \le \gamma_k \le \gamma_k^{\max},
	\end{gathered}
	\label{Eq_37}
\end{equation}
where $\gamma_k^{\min}$ and $\gamma_k^{\max}$ are the minimum and the maximum values of $\gamma_k$ allowed in the range of Eq. \eqref{Eq_33}, respectively. Therefore, all values of $\gamma_k$ on the interval $\left[\gamma_k^{\min},\gamma_k^{\max}\right]$ are accepted. In the present study, the value of $\gamma_k$ is selected to be $\gamma_k^{\min}$. For example, the values of $\gamma_k$ are able to exist in the interval $\left[0,1\right]$ if all values $m_k$ are available in the interval $\left[m_k^{\min},m_k^{\max}\right]$. Accordingly, all values of $\gamma_k$ become 0 by selecting the values of $\gamma_k^{\min}$, leading to identical results to those obtained with the minmod flux limiter. 

\begin{figure}
	\centering
	\includegraphics[width=\textwidth]{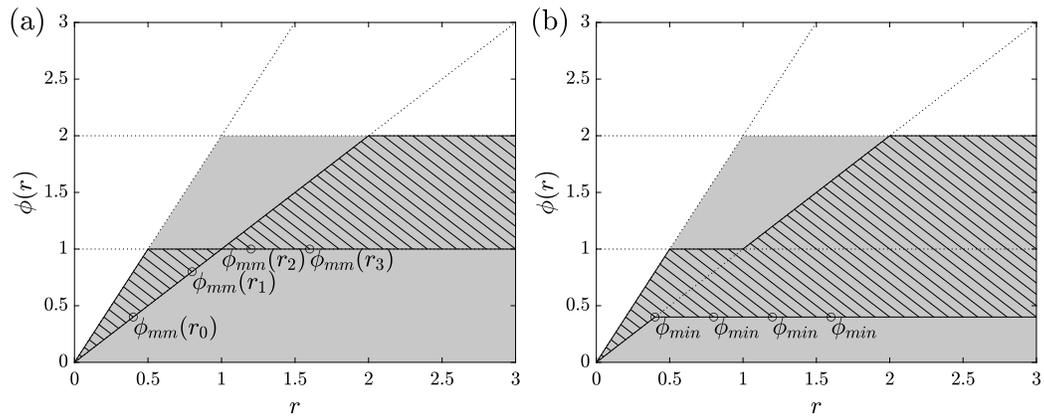}
	\caption{Flux limiter values obtained by applying (a) the minmod flux limiter and (b) the equal flux limiter.}
	\label{Fig_4}
\end{figure}

If the conditions for existence in Eqs. \eqref{Eq_35} and \eqref{Eq_35} are not satisfied, based on the equal flux limiter \citep{shiea2020}, the lower boundary of the variable flux limiter $\phi_{lower}$ in Eq. \eqref{Eq_25} is modified to equalize the values of the flux limiter for $(m_0,\ldots,m_3)^\trans$ into a minimum limiter value as follows:
\begin{equation}
	\phi_{lower}(r)=\max\left[0,\min\left(\phi_{\min},r\right)\right],
	\label{Eq_38}
\end{equation}
where $\phi_{\min} = \min \left[\phi_{mm}(r_0),\ldots,\phi_{mm}(r_3)\right]$. Then, the searchable region is expanded to allow the identical limiter value $\phi_{\min}$ for $(m_0,\ldots,m_3)^\trans$, as shown in Fig. \ref{Fig_4}(b). Thus, the reconstructed moment set $\mathbf{m}_{3}$ is represented as the convex combination of the moment sets in the upwind and downwind neighboring cells, $\mathbf{m}_{3,U}$ and $\mathbf{m}_{3,D}$.
\begin{equation}
	\mathbf{m}_{3} = \left(1-\lambda_{e}\phi_{\min}\right)\mathbf{m}_{3,U} + \lambda_{e}\phi_{\min}\mathbf{m}_{3,D}.
	\label{Eq_39}
\end{equation}
Consequently, the realizability and the TVD property are ensured for the reconstructed moment set $\mathbf{m}_{3}$, but the order of accuracy is partially reduced. 

Now consider the reconstruction of $m_4$ and $m_5$. First, let us denote the $k$th-order moment that belongs to the boundary causing $\Delta_k=0$ as $m_k^{-}$. Then, $m_k^{-}$ becomes the lower limit of the allowed values for $m_k$, and $m_k^{-}$ can be represented as a function of $\zeta$ in Eq. \eqref{Eq_14}:
\begin{equation}
	m_k^{-}=m_k(\zeta_0,\ldots,\zeta_{k-1},\;\zeta_k = 0).
	\label{Eq_40}
\end{equation}
The reconstructed value of $m_k$ is forced to be $m_k^{-}$ if $\Delta_{k-1}=0$. On the other hand, if $\Delta_{k-1}>0$, the reconstructed moment $m_k$ is determined to have the value at the lower boundary of the variable flux limiter (i.e., $\gamma_{k,e}=0$ in Eq. \eqref{Eq_26}). However, the value should not be lower than $m_k^{-}$. Thus, $m_k$ is bounded. 
\begin{equation}
	m_k = \left\{
	\begin{array}{cc}
	\max\left[m_{k,e}^{var}\left(\gamma_{k,e}=0\right),m_k^{-}\right] & \textmd{if}\quad \Delta_{k-1} > 0 \\
	m_k^{-} & \textmd{if}\quad \Delta_{k-1} = 0 \\
	\end{array}, \quad \textmd{for} \quad k=4,5.
	\right.
	\label{Eq_41}
\end{equation}
The moments obtained in Eq. \eqref{Eq_41} are realizable, but they have a possibility of losing the second-order accuracy and the TVD property. The overall reconstruction process is provided in Algorithm \ref{alg_2}.

After the moment sets are reconstructed at the cell faces, Algorithm \ref{alg_1} is applied to determine the time step size explicitly. The CFL number is defined as an upper limit of the time step size $\Delta t$ in Eq. \eqref{Eq_23}. Therefore, $\Delta t$ is computed as follows: 
\begin{equation}
	\max_{i}\left[\max_{eo}\left(\frac{1}{\Delta V_{i}}\mathbf{u}_{eo}\cdot{\mathbf{S}}_{eo}\right)\right]\Delta t = \min\left[\min_{i}\left({\frac{1}{1+N_{eo}}}\right),\textmd{CFL}\right].
	\label{Eq_42}
\end{equation}
Accordingly, the arbitrary moment set $\mathbf{m}_N^*$ defined in Eq. \eqref{Eq_18} is reformulated including the CFL number as follows:
\begin{equation}
	\mathbf{m}_{N}^{*}=\max\left[\left(1+N_{eo}\right),\textmd{CFL}^{-1}\right]\mathbf{m}_{N,i}^{n}-\sum_{eo=1}^{N_{eo}}\mathbf{m}_{N,eo}^{n},
	\label{Eq_43}
\end{equation}
In Eq. \eqref{Eq_43}, the coefficient before $\mathbf{m}_{N,i}^{n}$ increases as the CFL number decreases, resulting in $\mathbf{m}_{N}^{*}$ easily satisfying the realizability condition. Therefore, the situation that the modified moment set has severely deviated from the initially reconstructed moment set can be avoided by reducing the CFL number.

\section{Verification and discussion} \label{Ch4}
Three configurations are considered to evaluate the performance of the proposed variable flux limiter scheme. The first configuration comprises the pure advection Riemann problem in an 1D domain with a constant velocity. The second configuration is the case in which non-uniformly distributed moments are transported by a constant velocity in an 1D periodic domain. In the third configuration, moments are transported in a 2D domain with a steady Taylor-Green velocity field. For all three configurations, transport of the fifth-order moment set $\mathbf{m}_{5}=(m_0,\ldots,m_5)^\trans$ is considered. 

The results are compared with those obtained using the $\zeta$ simplified scheme \citep{laurent2017, passalacqua2020} and the equal limiter scheme \citep{shiea2020}. Algorithm \ref{alg_1} is applied for all reconstructed moment sets to explicitly obtain a time step size. As a time marching method, the second-order strong stability-preserving explicit Runge-Kutta method \citep{gottlieb2001} is applied for all schemes.

\subsection{1D Riemann problem} \label{Ch4-1}
\begin{figure}
	\centering
	\includegraphics[width=0.5\textwidth]{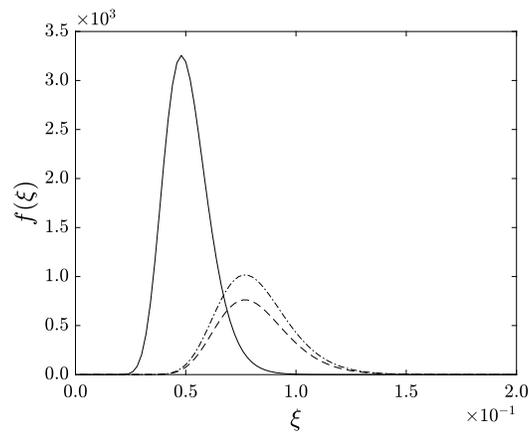}
	\caption{Number density functions (NDFs) of the boundary condition (\linesolid, $BC$) and the initial conditions (\linedotdash, $IC_1$; \linedash, $IC_2$).}
	\label{Fig_5}
\end{figure}

Results on the pure advection Riemann problem are examined in this section. The test domain is set in an 1D domain, and the spatial coordinate $x$ is defined on $[0,1]$. The spatial domain is discretized into 100 cells with the spacing $\Delta x$ of 0.01. The advection velocity $u$ is constant and set to $1.0$, and the CFL number is fixed to be 0.3. The boundary and initial conditions are defined using the log-normal distribution $f$ as follows:
\begin{equation}
	f(m_0,\mu,\sigma, \xi) = \frac{m_0}{\xi\sigma\sqrt{2\pi}}\exp\left(-\frac{(\ln{\xi}-\mu)^2}{2\sigma^2}\right),
\end{equation}
where $m_0$ is the zeroth-order moment, $\mu$ is the mean of the normal distribution, $\sigma$ is the standard deviation of the normal distribution, and $\xi$ is the size of the elements. Accordingly, the moments are defined as follows: 
\begin{equation}	
	m_{k}(m_0,\mu,\sigma)=\int_{0}^{+\infty} \xi^{k} f(m_0,\mu,\sigma, \xi) \mathrm{d} \xi.	
\end{equation}
One boundary condition ($BC$) and two initial conditions ($IC_1, IC_2$) are defined as follows:
\begin{equation}
	\begin{gathered}
	BC:\quad m_0=80,\quad \mu=\ln(0.05),\quad \sigma=0.2, \\
	IC_1:\quad m_0=40,\quad \mu=\ln(0.08),\quad \sigma=0.2, \\
	IC_2:\quad m_0=30,\quad \mu=\ln(0.08),\quad \sigma=0.2.
	\end{gathered}
\end{equation}
$IC_1$ and $IC_2$ are different only in $m_0$, which means the total number of elements. The NDFs of the boundary and the initial conditions are plotted in Fig. \ref{Fig_5}, and the corresponding moment sets are denoted as  $\mathbf{m}_{BC}$, $\mathbf{m}_{IC_1}$, and $\mathbf{m}_{IC_2}$, respectively. The boundary condition of $\mathbf{m}_{BC}$ is imposed on the left boundary, and the internal cells are initialized to have $\mathbf{m}_{IC_1}$ and $\mathbf{m}_{IC_2}$ for two different test cases, respectively.

\begin{figure}
	\centering
	\includegraphics[width=\textwidth]{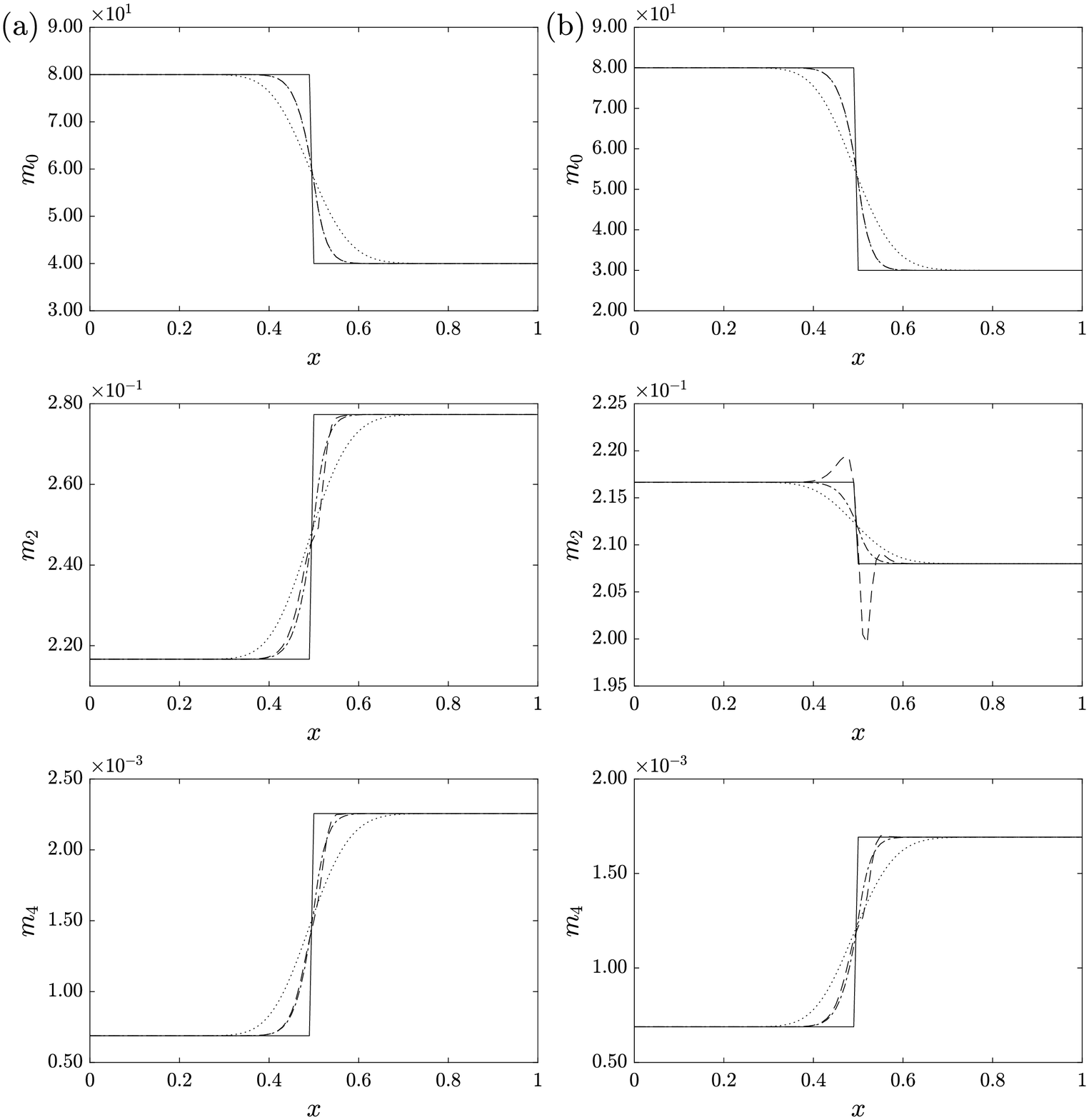}
	\caption{Spatial ditributions of the moments $m_0$, $m_2$, and $m_4$ for (a) the case using $BC$ and $IC_1$ and (b) the case using $BC$ and $IC_2$: \linesolid, analytical solution; \linedot, first-order upwind scheme; \linedash, $\zeta$ simplified scheme; \linedotdash, variable flux limiter scheme.}
	\label{Fig_6}
\end{figure}

Fig. \ref{Fig_6} shows the computational results obtained at $t=0.5$, indicating that the variable flux limiter scheme and the equal flux limiter scheme provide identical results. Thus, only the results of the variable flux limiter scheme are plotted in the figures. The reason why the two schemes provide identical results is discussed below. Let us assume the moment set of the cell $i$ is the convex combination of two moment sets, $\mathbf{m}_{BC}$ and $\mathbf{m}_{IC}$:
\begin{equation}
	\mathbf{m}_{i} = \alpha_{i}\mathbf{m}_{BC} + \left(1-\alpha_{i}\right)\mathbf{m}_{IC}.
\end{equation}
Then, the ratio of successive gradients $r_{k,e}$ presented in Eq. \eqref{Eq_7} is calculated as 
\begin{equation}
	r_{k,e} = \frac{m_{k,i}-m_{k,i-1}}{m_{k,i+1}-m_{k,i}}=\frac{\alpha_{i}-\alpha_{i-1}}{\alpha_{i+1}-\alpha_{i}}.
\end{equation}
Consequently, the value of the flux limiter $\phi(r_{k,e})$ is independent of the moment order $k$. Therefore, the reconstructed moment set at the face $e$ can be represented as the convex combination of $\mathbf{m}_{BC}$ and $\mathbf{m}_{IC}$ as follows:
\begin{equation}
	\mathbf{m}_{e} = \left(1-\frac{1}{2}\phi(r_e)\right)\mathbf{m}_i + \frac{1}{2}\phi(r_e)\mathbf{m}_{i+1}=\alpha_{e}\mathbf{m}_{BC} + \left(1-\alpha_{e}\right)\mathbf{m}_{IC}. 
\end{equation}
As a result, during computation, the moment sets on the domain were preserved, forming the convex combination of $\mathbf{m}_{BC}$ and $\mathbf{m}_{IC}$. Here, the equal flux limiter is identical to the minmod flux limiter because the value of $\phi(r_{k,e})$ is the same for all moment orders. In addition, the variable flux limiter converges to the minmod flux limiter as much as possible. Thus, there are two schemes that provide identical results.

\begin{figure}
	\centering
	\includegraphics[width=\textwidth]{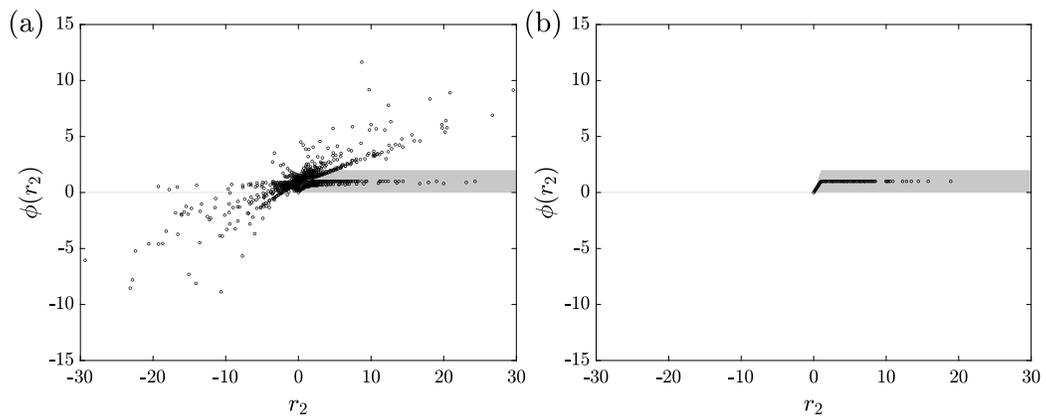}
	\caption{Distributions of the flux limiter values in the TVD diagram for $m_2$ in the case $IC_2$  with (a) the $\zeta$ simplified scheme and (b) the variable flux limiter scheme. The shaded area refers to the TVD region.}
	\label{Fig_7}
\end{figure}

Let us now return to the numerical results in Fig. \ref{Fig_6}. In the result of the second-order moment $m_2$ for the case using $BC$ and $IC_2$, the $\zeta$ simplified scheme causes noticeable overshoot and undershoot. This phenomenon indicates that the $\zeta$ simplified scheme does not guarantee the TVD characteristic. In the $\zeta$ simplified scheme, the minmod flux limiter is applied on the $\zeta$ quantities and not on the moments. Hence, the TVD characteristic is not guaranteed except for $m_0$, which is identical to $\zeta_0$. Fig. \ref{Fig_7} shows distributions of flux limiter values that are reversely calculated from the reconstructed moment values for $m_2$ in the test case employing $IC_2$, and all values obtained during the computation are displayed. As shown in Fig. \ref{Fig_7}(a), a significant number of flux limiter values obtained using the $\zeta$ simplified scheme appear to deviate from the TVD region. On the other hand, the values of the variable flux limiter are the same as when using the minmod flux limiter.

\subsection{1D non-uniform initial distribution} \label{Ch4-2}
In the second configuration, moments are initially distributed differently depending on their locations. The test domain is set in an 1D domain, and the periodic boundary condition is imposed. The spatial coordinate $x$ is defined on $[0,1]$. The advection velocity $u$ is constant and set to $1.0$, and the CFL number is fixed to be 0.3. The spatial domain is discretized into varying number of cells from $50$ to $3200$ to evaluate the accuracy of the schemes. The following three initial conditions in \citep{laurent2017, passalacqua2020} which are denoted as the regular initial NDF, the oscillating initial $\zeta$, and the multi-modal initial NDF, respectively, are considered:

Regular initial NDF:
\begin{equation}
	\begin{gathered}
	m_k(x) = 16x^2(1-x)^2\frac{\beta\left(\lambda(x)+k, \mu(x)\right)}{\beta\left(\lambda(x), \mu(x)\right)}, \\
	\lambda(x)=\frac{7}{2}+\frac{3}{2}\sin\left(2\pi x\right), \\ 
	\mu(x)=\frac{7}{2}-\frac{3}{2}\cos\left(2\pi x\right). \\
	\end{gathered}
\end{equation}

Oscillating initial $\zeta$:
\begin{equation}
	\begin{gathered}
	m_k(x) = m_k\left(\zeta_0(x),\ldots,\zeta_k(x)\right), \\
	\zeta_0(x) = 16x^2(1-x)^2, \\
    \zeta_k(x) = \frac{x}{2}\left[1.01+\cos\left(\frac{\pi k}{2}x\right)\right]. \\
	\end{gathered}
\end{equation}

Multi-modal initial NDF:
\begin{equation}
	\begin{gathered}
	m_k(x) = w_1(x)\xi_1^k + w_2(x)\xi_2^k + w_3(x)\lambda^k(x)\Gamma\left(1+\frac{k}{\kappa(x)}\right), \\
	w_1(x) = \left\{
	\begin{array}{cc}
	16x^2(1-x)^2 & \quad 0 \le x \le 1 \\
	0 & \quad \textmd{otherwise} \\
	\end{array},
	\right. \\
	w_2(x) = \left\{
	\begin{array}{cc}
	\frac{256}{81}(4 x-1)^{2}(1-x)^{2} & \frac{1}{4} \le x \le 1 \\
	0 & \textmd{otherwise} \\
	\end{array},
	\right. \\
	w_3(x) = \left\{
	\begin{array}{cc}
	9(3 x-1)^{2}(1-x)^{2} & \frac{1}{3} \le x \le 1 \\
	0 & \textmd{otherwise} \\
	\end{array},
	\right. \\
	\lambda(x) = \left\{
	\begin{array}{cc}
	\lambda_{\min} = 0.02 & 0 \le x \le \frac{1}{3} \\
	\lambda_{\min}(2-3x)^2(6x-1) + \lambda_{\max}(3x-1)^2(5-6x) & \frac{1}{3} < x \le \frac{2}{3} \\
	\lambda_{\max} = 0.7 & \frac{2}{3} < x \le 1 
	\end{array},
	\right. \\
	\kappa(x) = \left\{
	\begin{array}{cc}
	\kappa_{\min} = 3 & 0 \le x \le \frac{1}{3} \\
	\kappa_{\min}(2-3x)^2(6x-1) + \kappa_{\max}(3x-1)^2(5-6x) & \frac{1}{3} < x \le \frac{2}{3} \\
	\kappa_{\max} = 10 & \frac{2}{3} < x \le 1 
	\end{array},
	\right. \\	
	\end{gathered}
\end{equation}
where $\xi_1 = 0.02$ and $\xi_2 = 2\xi_1 = 0.04$. 

The regular initial NDF is defined with the beta distribution, and the corresponding moments are in the interior of the moment space, which is far from the boundary. The oscillating initial $\zeta$ is defined using the $\zeta$ quantities oscillating at different frequencies, and the $\zeta$ values are designed to be close to 0 at their local minimum points. At these points, the corresponding moments are in the interior close to the boundary of the moment space. The multi-modal initial NDF is defined to have a different number of quadrature depending on the location. The multi-modal initial NDF is represented with only one weighted Dirac delta function centered on the quadrature abscissa $\xi_1=0.02$ for $x\in[0, 1/4]$ and two weighted Dirac delta functions centered on $\xi_1=0.02$ and $\xi_2=0.04$ for $x\in[1/4, 1/3]$. This distribution is encountered in practical problems containing nucleation and aggregation of particles. Therefore, the corresponding moments are on the boundary of the moment space for $x\in[0, 1/3]$.

\begin{figure}
	\centering
	\includegraphics[width=\textwidth]{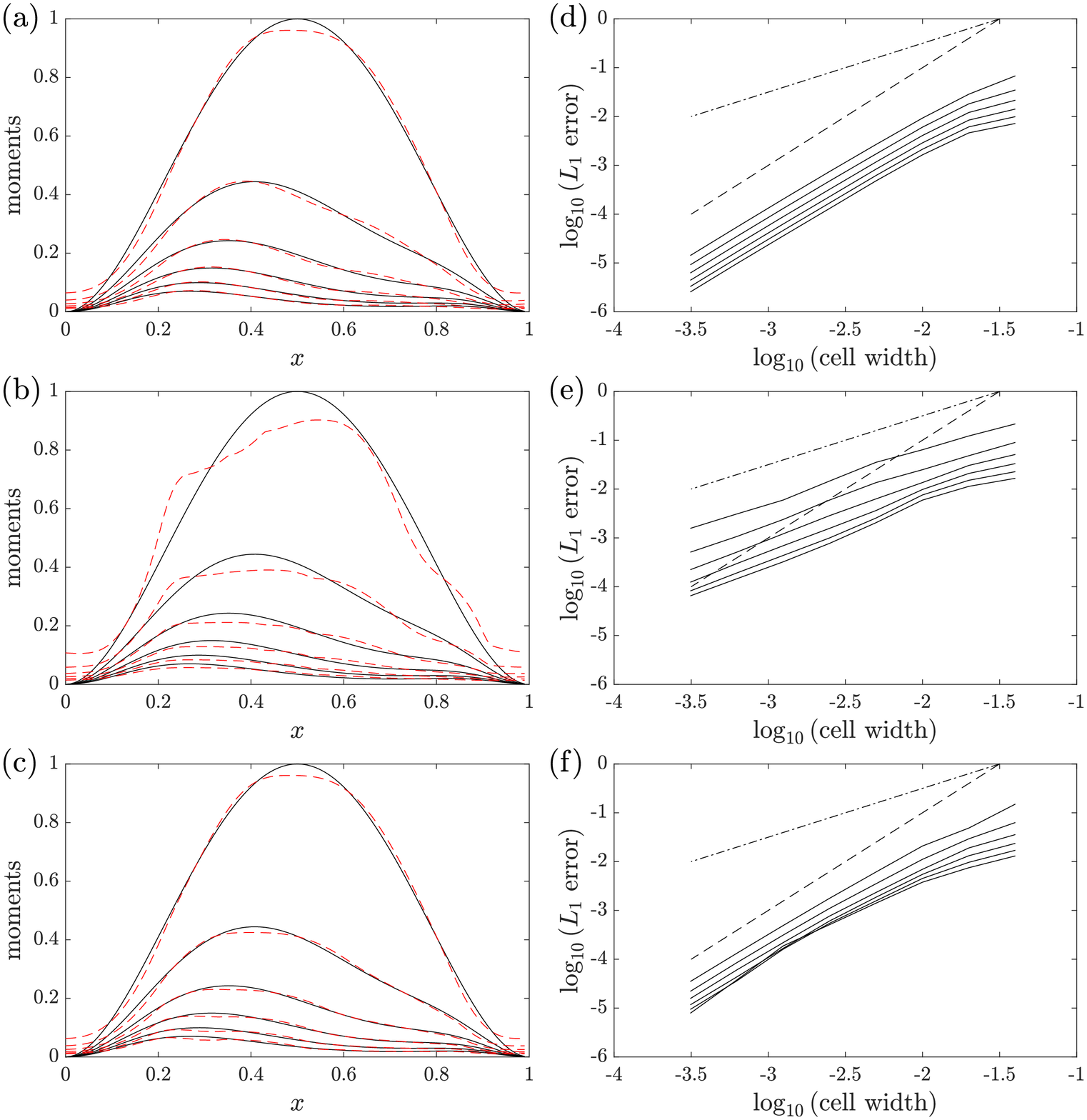}
	\caption{Comparison of results for the regular initial NDF. (a, b, c) Spatial ditributions of the moments of order 0 to 5 (top to bottom lines) for the analytical solutions (\linesolid) and for the simulation results (\textcolor{red}{\linedash}), and (d, e, f) $L_1$ norms of errors on the moments of order 0 to 5 (top to bottom lines) for the simulations results (\linesolid) with the first-order slope (\linedotdash) and the second-order slope (\linedash) at $t=5$ obtained using three different schemes: (a, d) $\zeta$ simplified scheme; (b, e) equal flux limiter scheme; (c, f) variable flux limiter scheme.}
	\label{Fig_8}
\end{figure}

\begin{figure}
	\centering
	\includegraphics[width=\textwidth]{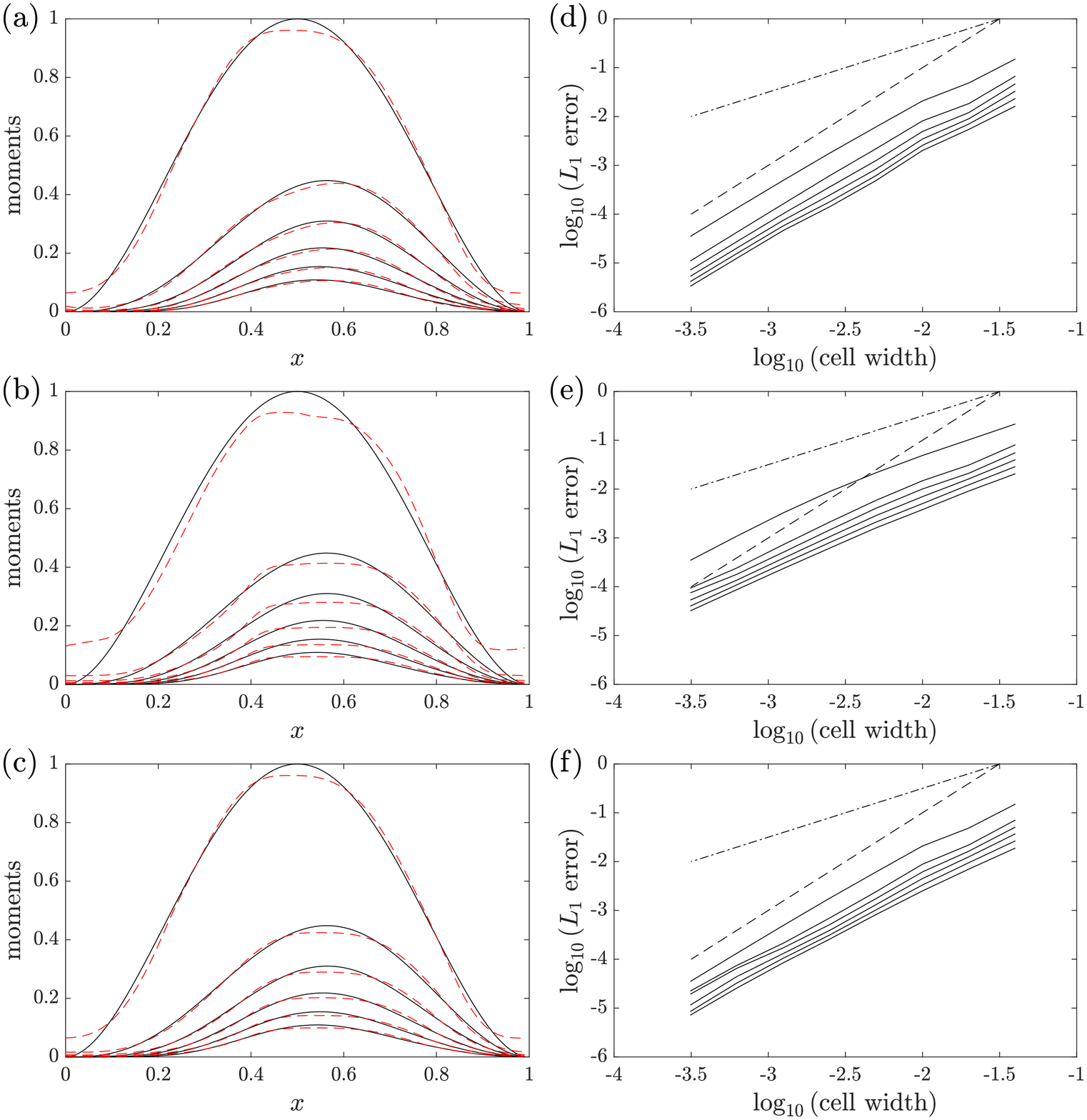}
	\caption{Comparison of results for the oscillating initial $\zeta$. (a, b, c) Spatial ditributions of the moments of order 0 to 5 (top to bottom lines) for the analytical solutions (\linesolid) and for the simulation results (\textcolor{red}{\linedash}), and (d, e, f) $L_1$ norms of errors on the moments of order 0 to 5 (top to bottom lines) for the simulations results (\linesolid) with the first-order slope (\linedotdash) and the second-order slope (\linedash) at $t=5$ obtained using three different schemes: (a, d) $\zeta$ simplified scheme; (b, e) equal flux limiter scheme; (c, f) variable flux limiter scheme.}
	\label{Fig_9}
\end{figure}

\begin{figure}
	\centering
	\includegraphics[width=\textwidth]{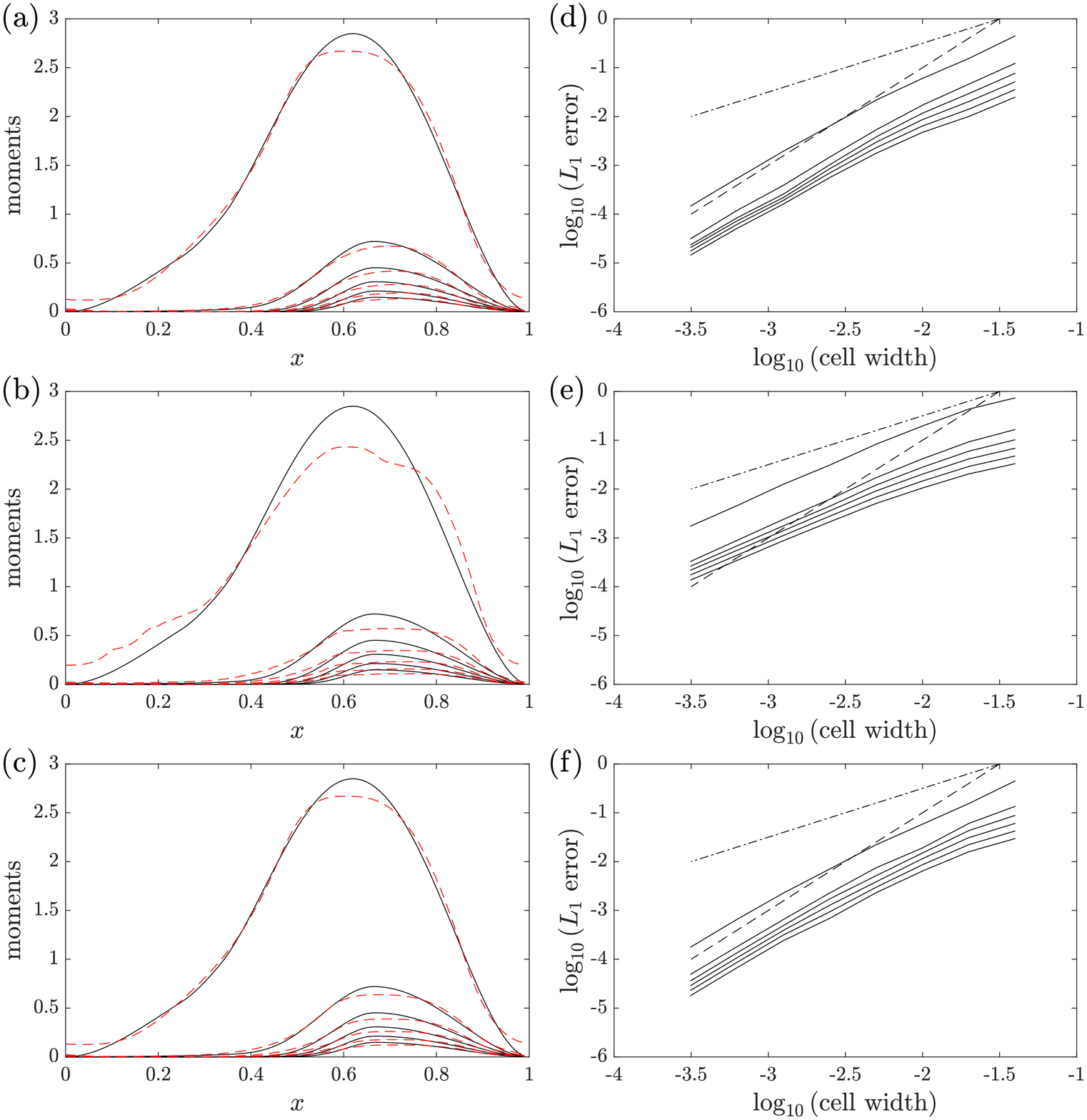}
	\caption{Comparison of results for the multi-modal initial NDF. (a, b, c) Spatial ditributions of the moments of order 0 to 5 (top to bottom lines) for the analytical solutions (\linesolid) and for the simulation results (\textcolor{red}{\linedash}), and (d, e, f) $L_1$ norms of errors on the moments of order 0 to 5 (top to bottom lines) for the simulations results (\linesolid) with the first-order slope (\linedotdash) and the second-order slope (\linedash) at $t=5$ obtained using three different schemes: (a, d) $\zeta$ simplified scheme; (b, e) equal flux limiter scheme; (c, f) variable flux limiter scheme.}	
	\label{Fig_10}
\end{figure}

Figs. \ref{Fig_8}--\ref{Fig_10} show spatial distributions of moments and $L_1$ norms of errors on the moments for the three initial conditions, respectively. The results are obtained at $t=5$ by employing three different schemes, including the $\zeta$ simplified scheme, the equal flux limiter scheme, and the variable flux limiter scheme. The equal flux limiter results show a relatively large error compared to the other two methods, as shown in Figs. \ref{Fig_8}--\ref{Fig_10}(b) and (e). Let us look at the result of the regular initial NDF condition that showing the most severe distortion (see Fig. \ref{Fig_8}(b) and (e)). In this case, the numerical order of accuracy worsens almost to the first order. 

\begin{figure}
	\centering
	\includegraphics[width=\textwidth]{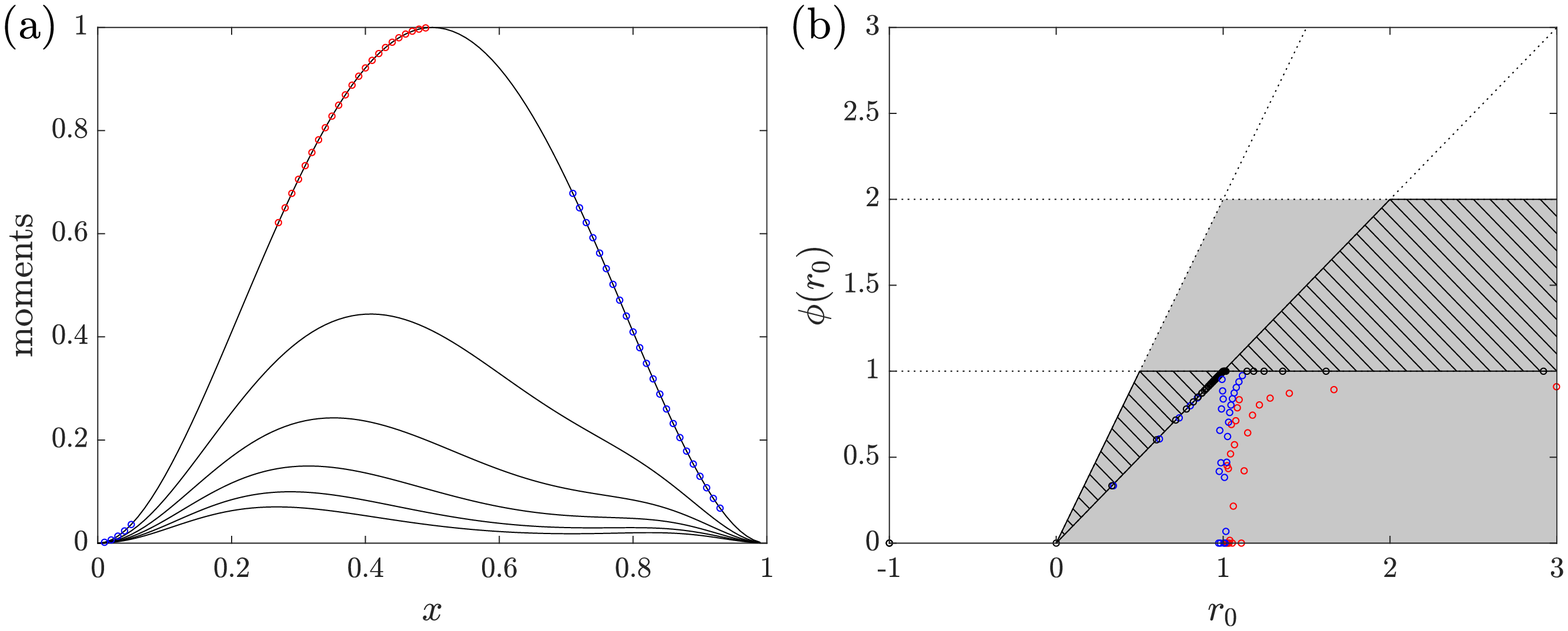}
	\caption{(a) Spatial distribution of the moments of order 0 to 5 (top to bottom lines), and (b) the flux limiter values for $m_0$ calculated by applying the equal flux limiter for the regular initial NDF. The points marked in red and blue in (b) are outside the second-order TVD region and corresponding locations are indicated with the same colors in (a).}
	\label{Fig_11}
\end{figure}

The reason why the accuracy of the equal flux limiter scheme shows a relatively large dependence on the initial condition can be identified on the TVD diagram. Fig. \ref{Fig_11} shows initial distributions of moments of the regular initial NDF and the flux limiter values for $m_0$ calculated using the equal flux limiter in these distribution. The values marked in red and blue are lower than those calculated by the minmod flux limiter (i.e., $\phi_e^{\min} < \phi_e^{mm}(r_{0,e})$), and the corresponding locations are also indicated with the same colors. The flux limiter values fall into the lower order of accuracy region at a considerable number of points, leading to the degenerated accuracy, as shown in Fig. \ref{Fig_11}. The described problem can be severe when the ratios of successive gradient $r_{k,e}$ have largely deviated from each other. In particular, in the worst-case scenario, the equal flux limiter value becomes 0 (i.e., $\phi_e^{\min}=0$) at the locations where the moments of the other orders have an extreme value in their distribution because the ratio becomes lower than 0 at the extremum (i.e., $r_{k,e}<0$). The slopes corresponding to the numerical orders of accuracy for $m_0$ and $m_3$ are listed in Table \ref{table_1}. Based on the table, the orders of accuracy greatly vary depending on the initial condition when the equal flux limiter is employed, which can be explained using the analysis on the TVD diagram. 

\begin{figure}
	\centering
	\includegraphics[width=\textwidth]{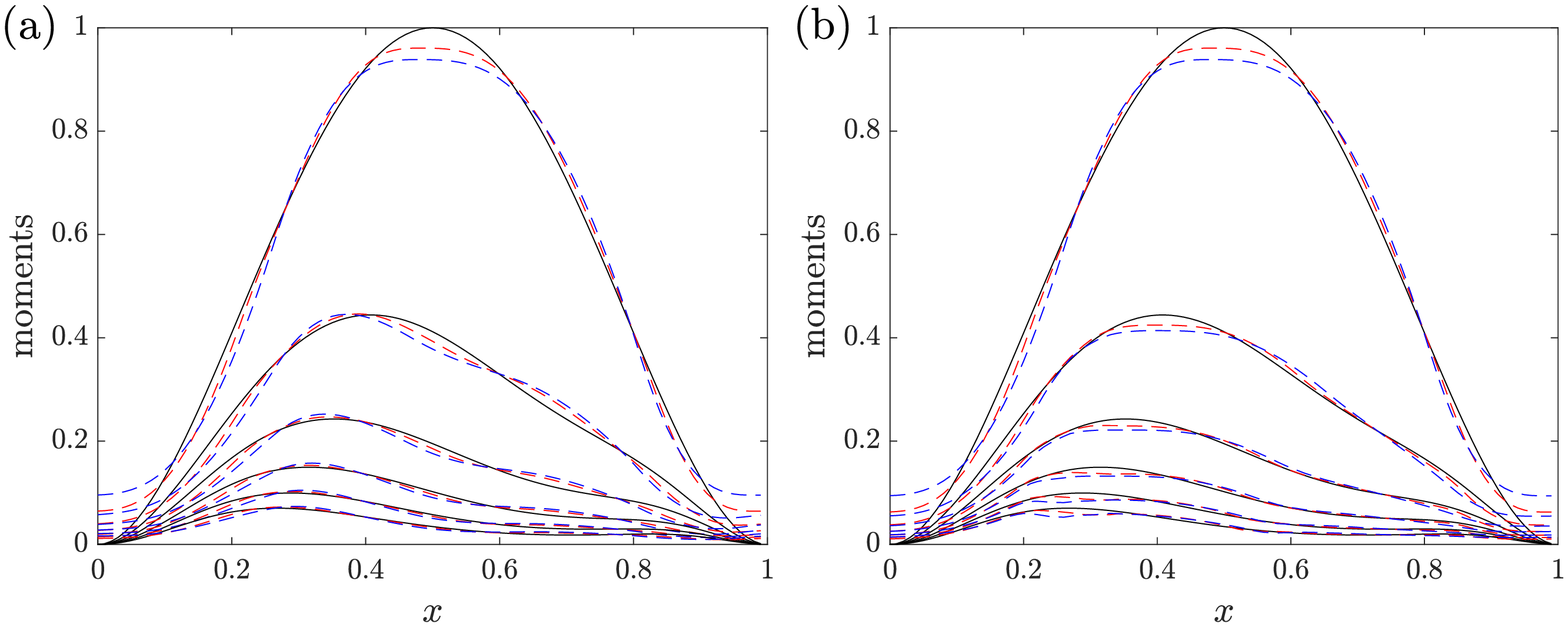}
	\caption{\textmd{Spatial ditributions of the moments of order 0 to 5 (top to bottom lines) for the analytical solutions (\linesolid) and for the simulation results at $t=5$ (\textcolor{red}{\linedash}) and $t=10$ (\textcolor{blue}{\linedash}) for the regular initial NDF: (a) $\zeta$ simplified scheme; (b) variable flux limiter scheme.}}
	\label{Fig_12}
\end{figure}

Let us now consider the boundedness of the moments. For the regular initial NDF condition, Fig. \ref{Fig_12} shows the computation results at $t=5$ (red dashed lines) and $t=10$ (blue dashed lines). These results are obtained by applying the $\zeta$ simplified scheme (see Fig. \ref{Fig_12}(a)) and the variable flux limiter (see Fig. \ref{Fig_12}(b)), respectively. The results provided by employing the $\zeta$ simplified scheme form the oscillatory solutions around the analytical solutions for the moments $m_k$ for $k>0$. Based on the comparison of the results at $t=5$ and $t=10$, the oscillations are slightly amplified. Therefore, the unbounded solutions were observed at $t=10$ for the moments $m_k$ for $k>0$. This is consistent with the analysis described in Section \ref{Ch4-1}, stating that the TVD property is guaranteed only for $m_0$ for the $\zeta$ simplified scheme. In the case of the variable flux limiter, the TVD property is guaranteed up to the third-order moment in the reconstruction step. Thus, the bounded solutions are provided for those moments, as shown in Fig \ref{Fig_12}(b). However, the TVD property is not always satisfied for $m_4$ and $m_5$ as described in Section \ref{Ch3}. Therefore, the oscillatory shapes are observed in their solutions. In terms of boundedness, the equal flux limiter scheme can efficiently preserve the TVD property for all moments but it has the problem that the accuracy decreases significantly when the difference in the shape of distribution between moments is severe.

\subsection{2D Taylor-Green vortex} \label{Ch4-3}
In the third configuration, transport of moments in a steady vortical velocity field is considered. The test domain is set in a 2D domain, and the spatial coordinates $(x,y)$ are defined on $[0,0.5]\times[0,0.5]$. The velocity components are defined as follows:
\begin{equation}
u_x(x,y) = \sin(2\pi x)\cos(2\pi y),\quad u_y(x,y) = -\cos(2\pi x)\sin(2\pi y).
\end{equation}
The CFL number is fixed to be 0.2, and the spatial domain is uniformly discretized into varying number of cells from $50\times50$ to $1600\times1600$. In the present study, the 2D regular initial NDF condition in \citep{passalacqua2020} is employed. This initial condition impose moments that exist inside the moment space away from the boundary. The moments are initially defined using the radial distance from the point $(1/8,1/8)$ as follows:
\begin{equation}
	\begin{gathered}
		L=8\sqrt{\left(x-\frac{1}{8}\right)^2+\left(y-\frac{1}{8}\right)^2}, \\
		\lambda(L)=\frac{7}{2}+\frac{3}{2}\sin\left(2\pi (1-L)\right), \\
		\mu(L)=\frac{7}{2}-\frac{3}{2}\cos\left(2\pi (1-L)\right), \\
		\theta(L)=\frac{1}{2}+\frac{1}{2}\tanh\left[\tan\left(\pi\left(\frac{1}{2}-L\right)\right)\right], \\
		m_k(L) = \frac{m_{k-1}(L)\left[\lambda(L)+k-1\right]}{\lambda(L)+\mu(L)+k-1}\theta(L),\quad k=1,\ldots,5,
	\end{gathered}
\end{equation}
where $m_0(L)=\theta(L)$. 

\begin{figure}
	\centering
	\includegraphics[width=0.47\textwidth]{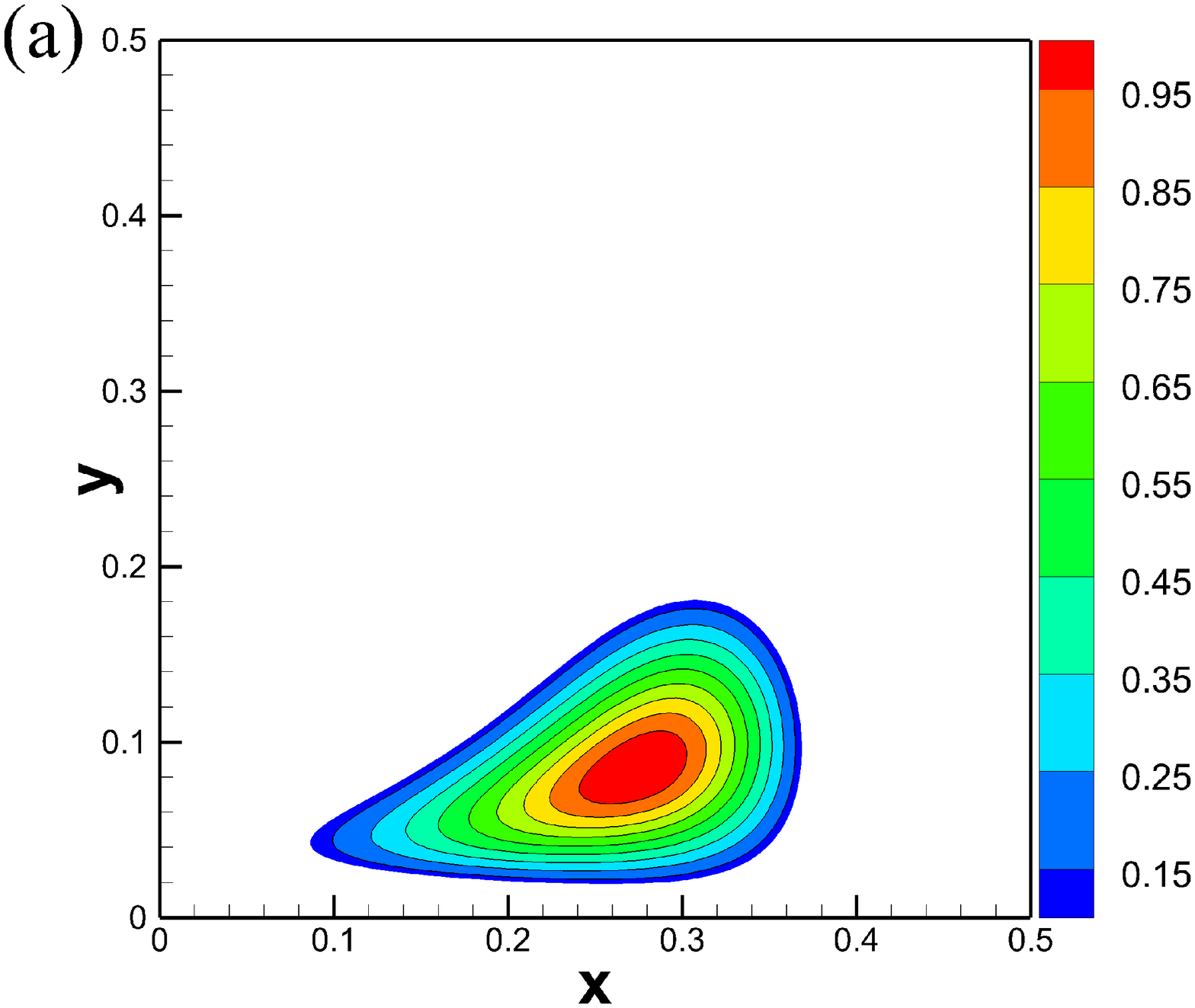}
	\includegraphics[width=0.47\textwidth]{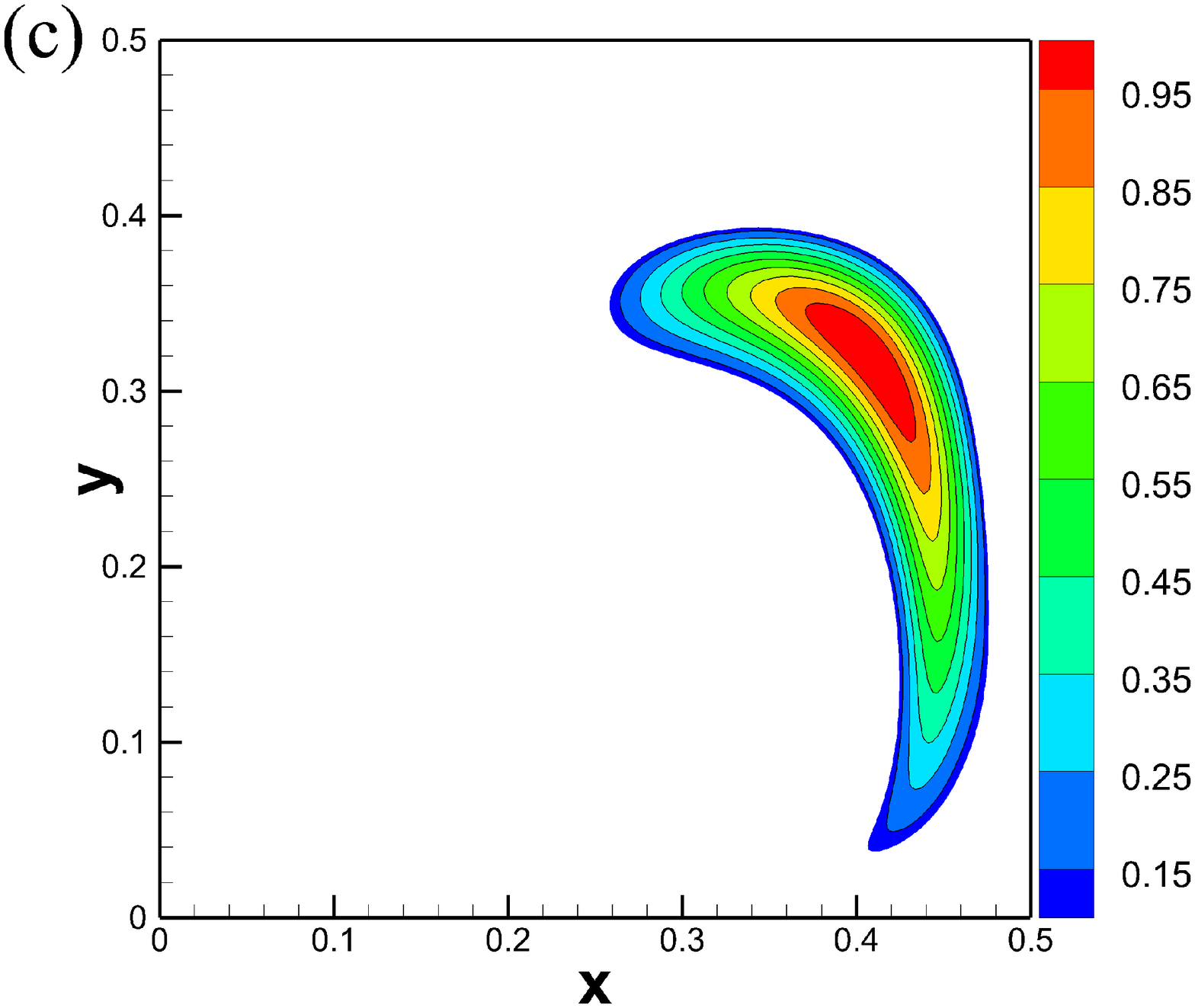}
	\includegraphics[width=0.47\textwidth]{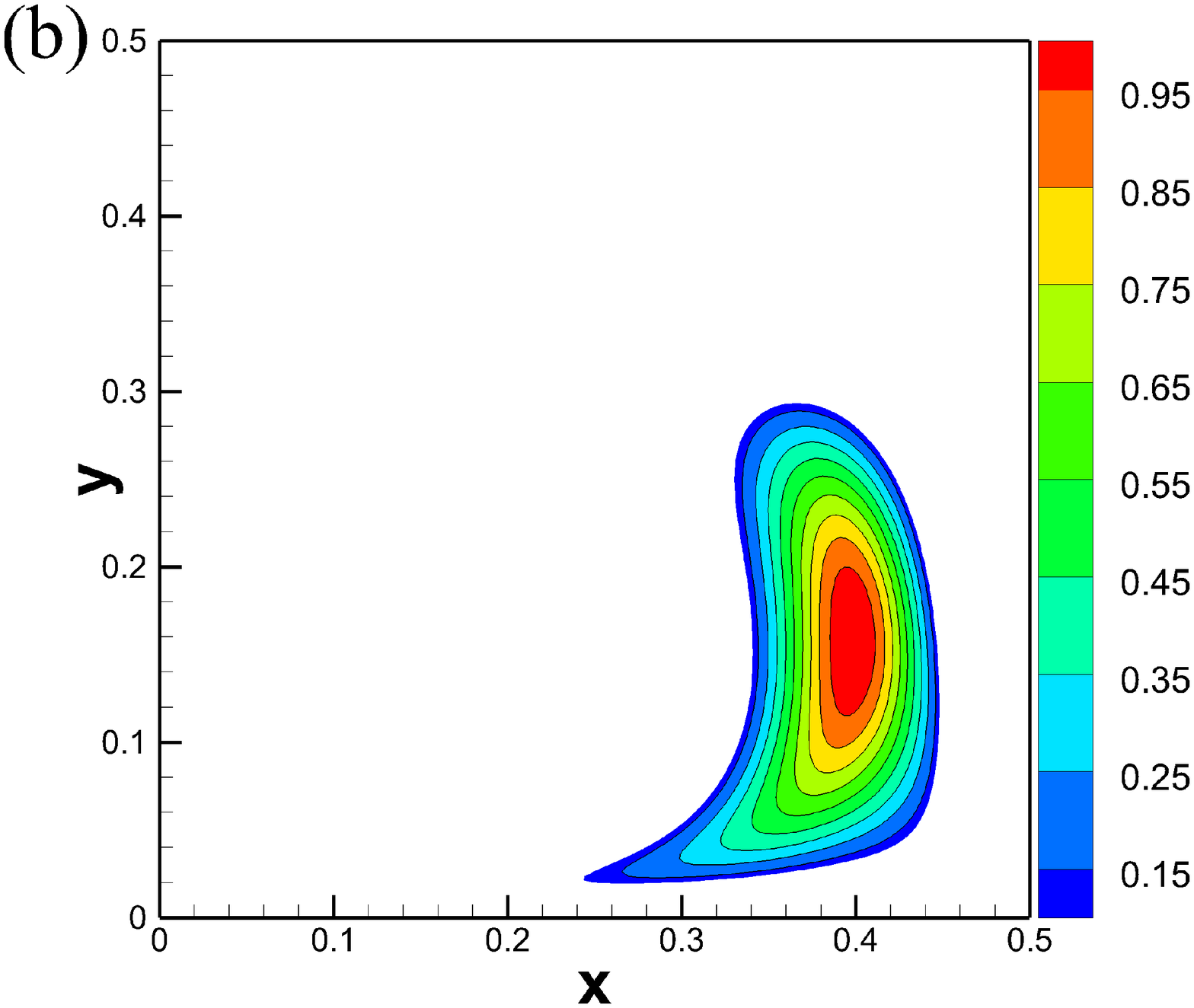}
	\includegraphics[width=0.47\textwidth]{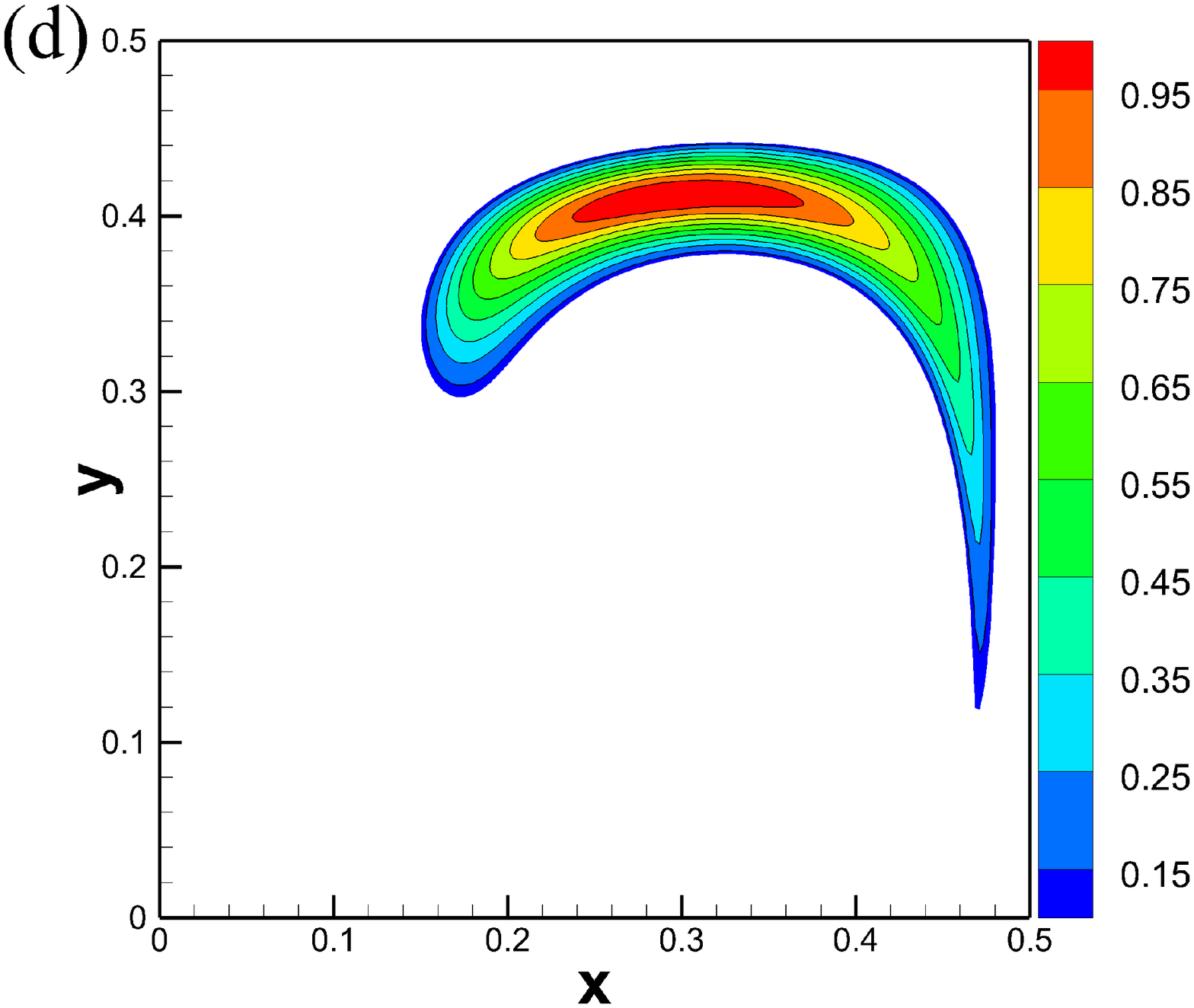}
	\caption{Spatial distributions of the zeroth-order moment $m_0$ in the reference solution at (a) $t=0.2$, (b) $t=0.4$, (c) $t=0.6$, and (d) $t=0.8$.}
	\label{Fig_13}
\end{figure}

A reference solution is obtained by solving reverse characteristics from each position using the fourth-order ODE solver because the value of the moments are conserved along the characteristic lines in the incompressible flow field. Spatial distributions of the zeroth-order moment in the computed reference solution are shown in Fig. \ref{Fig_13}.

\begin{figure}
	\centering
	\includegraphics[width=0.47\textwidth]{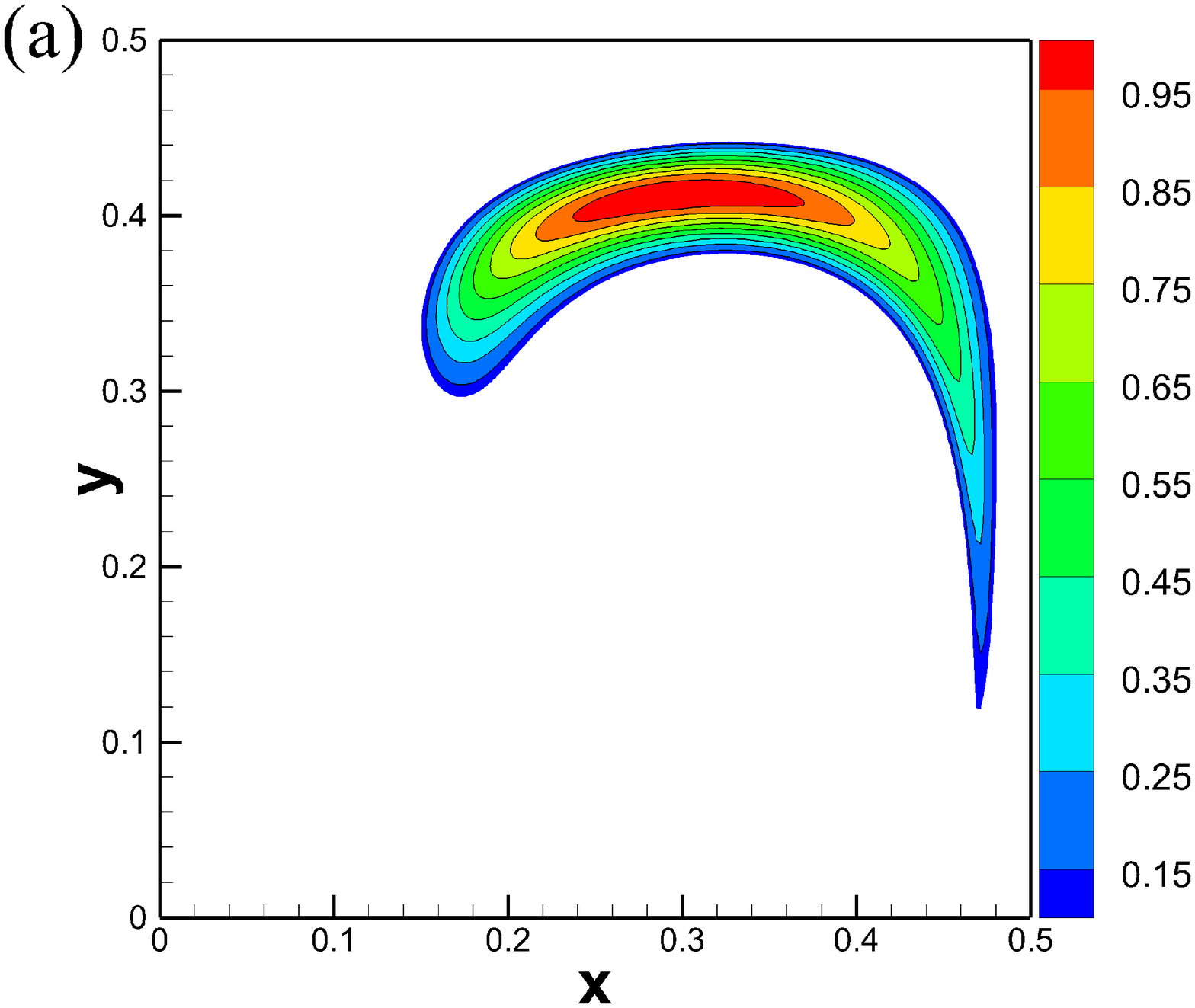}
	\includegraphics[width=0.47\textwidth]{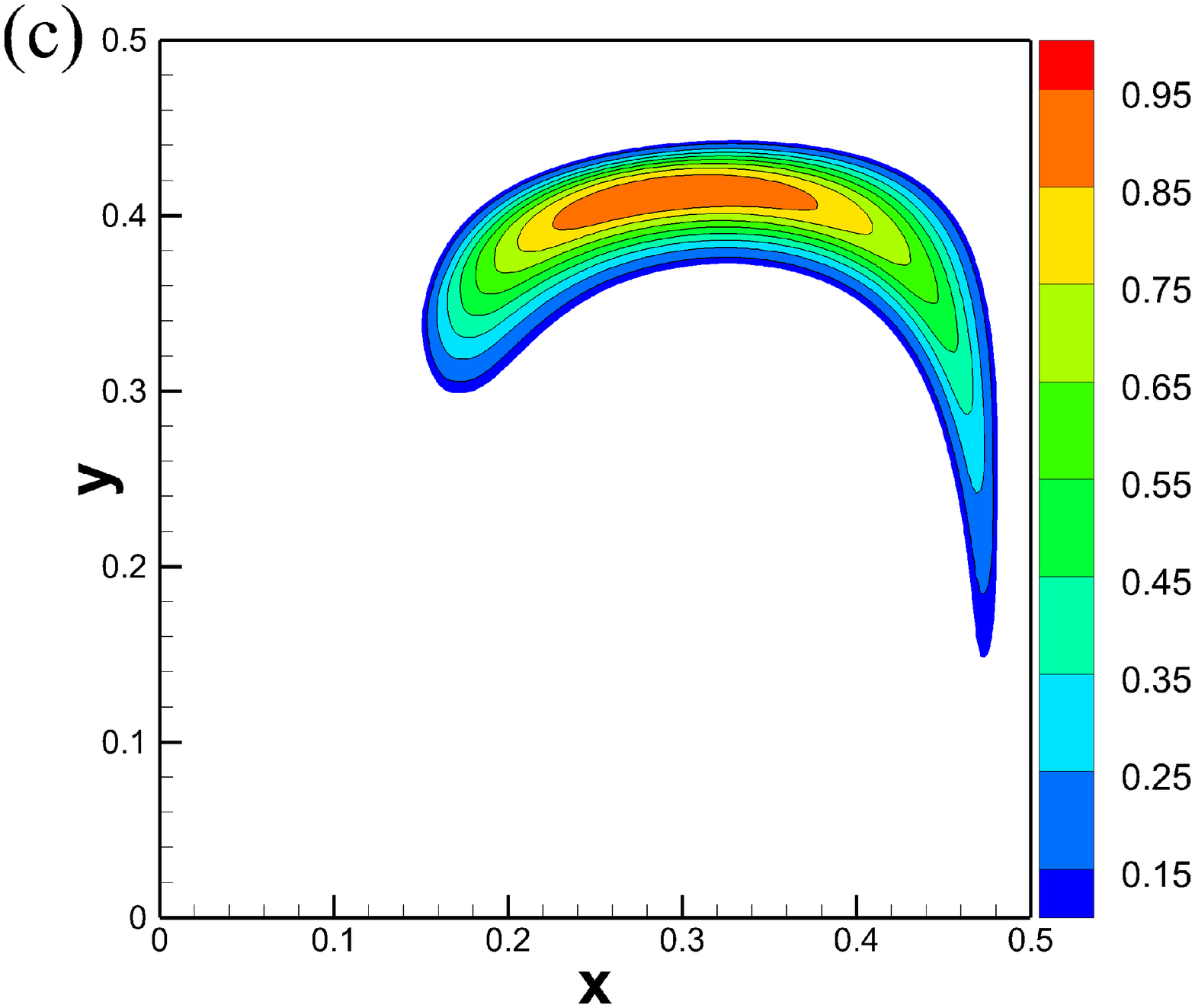}
	\includegraphics[width=0.47\textwidth]{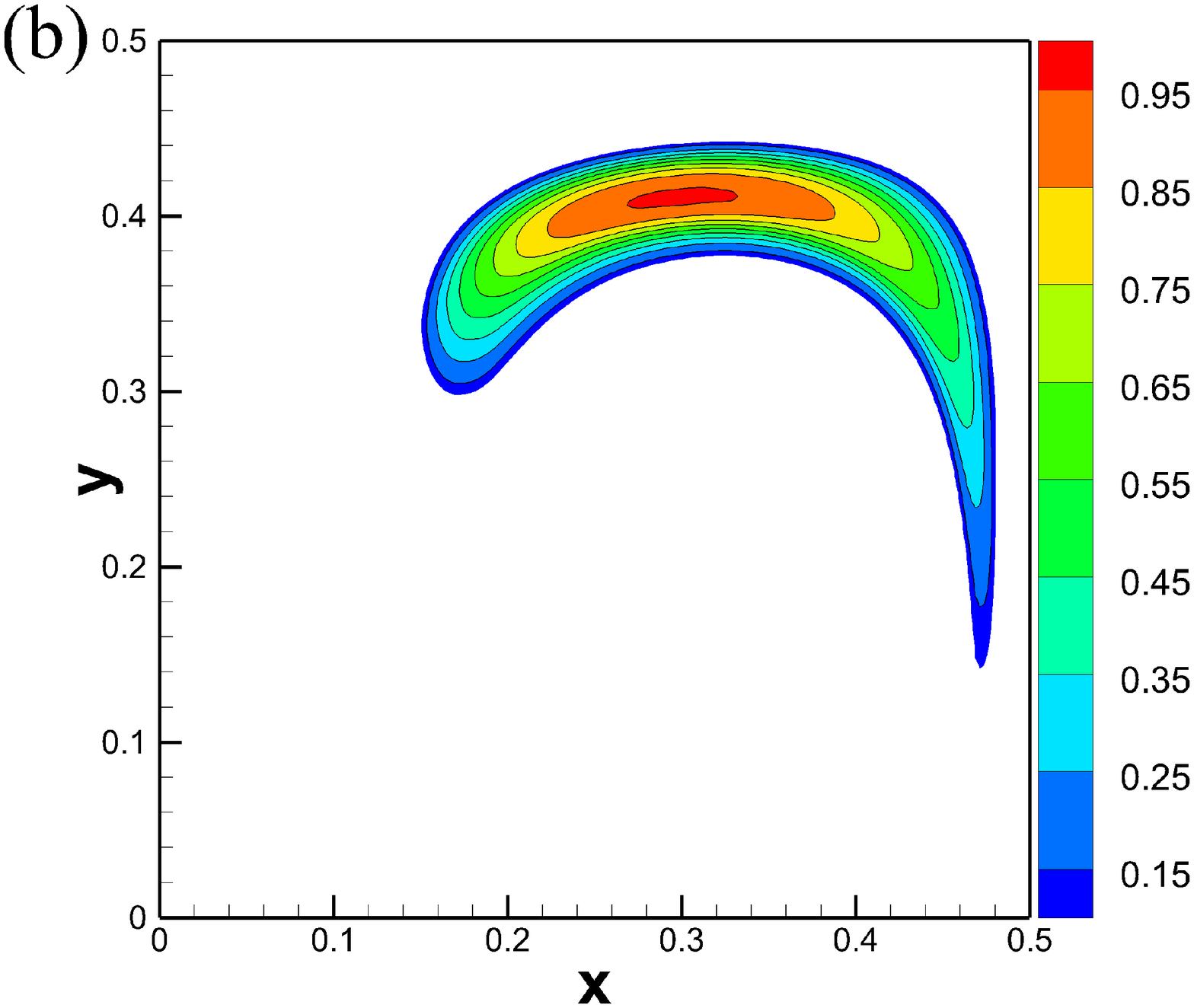}	
	\includegraphics[width=0.47\textwidth]{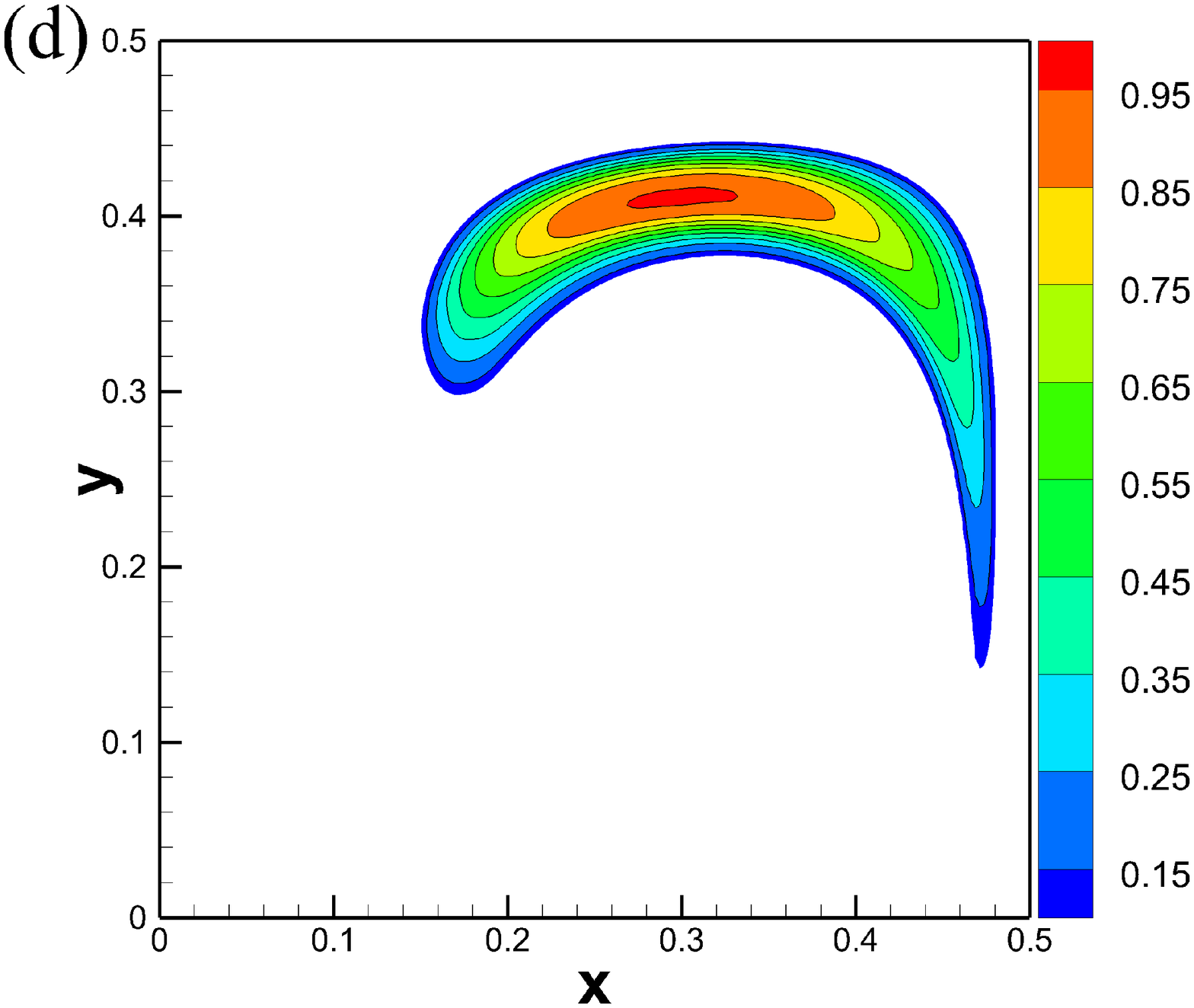}			
	\caption{Spatial distributions of the zeroth-order moment $m_0$ at $t=0.8$ on the $200\times200$ size mesh: (a) reference solution; (b) $\zeta$ simplified scheme; (c) equal flux limiter scheme; (d) variable flux limiter scheme.}
	\label{Fig_14}
\end{figure}

The distribution of $m_0$ transported using the variable flux limiter till $t=0.8$ is shown in Fig. \ref{Fig_14} with the results of the $\zeta$ simplified scheme and the equal flux limiter scheme. It is difficult to distinguish between the results obtained using the $\zeta$ simplified scheme and the variable flux limiter scheme (see Fig. \ref{Fig_14}(b) and (d)). However, the equal flux limiter scheme provides the more diffusive result (see Fig. \ref{Fig_14}(c)). 

\begin{figure}
	\centering
	\includegraphics[width=\textwidth]{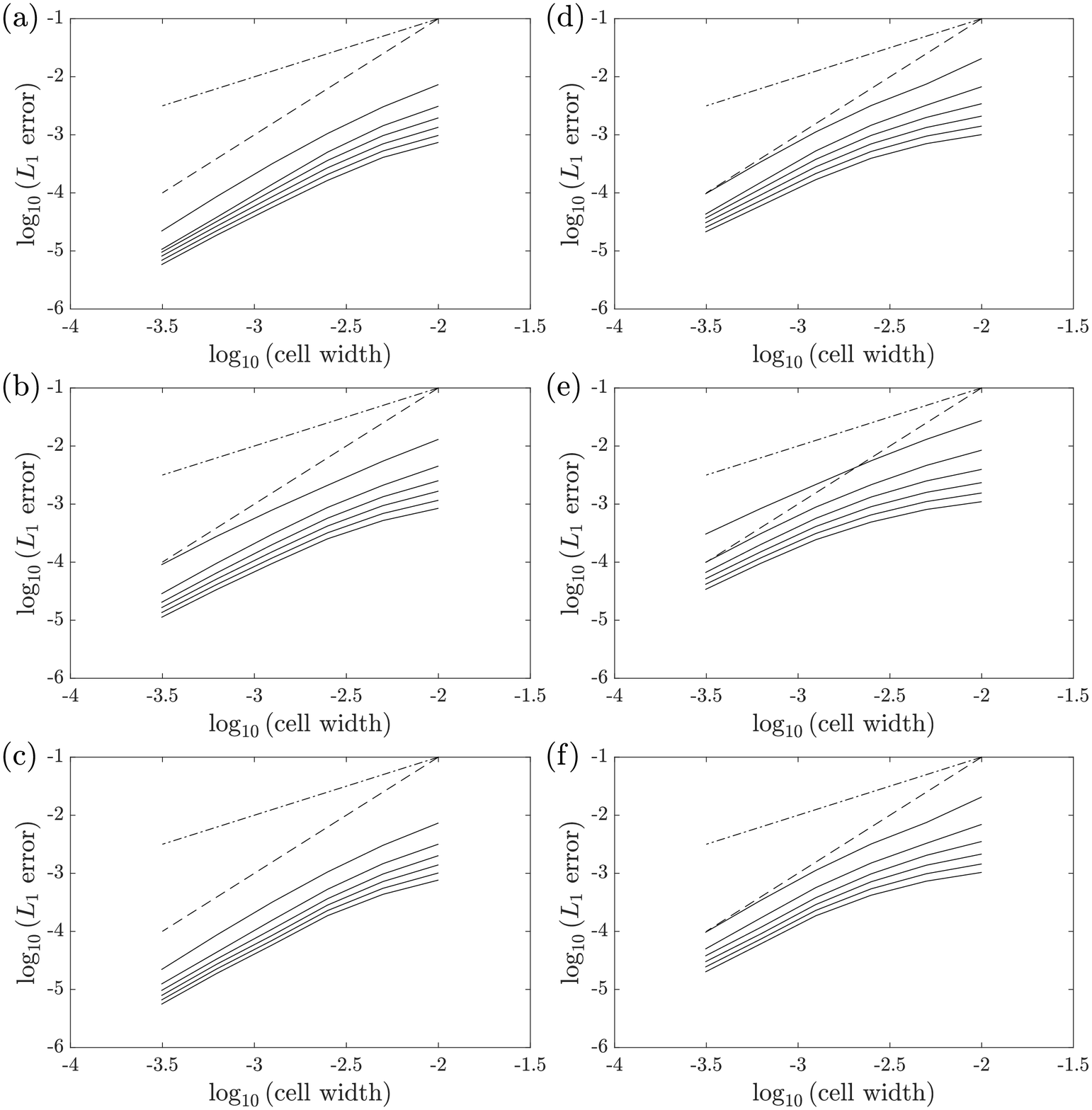}
	\caption{$L_1$ norms of errors on the moments of order 0 to 5 (top to bottom lines) for the simulation results (\linesolid) with the first-order slope (\linedotdash) and the second-order slope (\linedash) at (a, b, c) $t=0.4$ and (d, e, f) $t=0.8$ obtained using three different schemes: (a, d) $\zeta$ simplified scheme; (b, e) equal flux limiter scheme; (c, f) variable flux limiter scheme (e, f).}
	\label{Fig_15}
\end{figure}

The error is defined as a difference from the reference solution, and $L_1$ norms of errors are plotted as a function of the cell width at times $t=0.4$ and $t=0.8$ in Fig. \ref{Fig_15}. Table \ref{table_2} lists the slopes corresponding to the numerical order of accuracy for $m_0$ and $m_3$. The results obtained using the $\zeta$ simplified scheme and the variable flux limiter scheme show a similar order of accuracy, while the equal flux limiter scheme provides relatively worse results.

\section{Conclusion} 
In the present study, a new second-order TVD method with variable flux limiters, which is named as a variable flux limiter scheme, has been proposed to overcome the non-realizability issue that occurs while applying the conventional second-order TVD schemes to solve the moment transport equations. The underlying idea is to reconstruct the realizable moment set that satisfies the second-order TVD property at the cell face by allowing the flexible selection of the flux limiter values within the second-order TVD region. This idea has been realized by devising a variable flux limiter that combines two different flux limiters. 

The proposed method can satisfy the TVD property by applying the flux limiters directly to the reconstruction of the moments. At the same time, by allowing variable flux limiter values to each moment under the realizability condition, this method prevents the accuracy loss caused by equalizing the flux limiter values to ensure TVD characteristics. The conditions to simultaneously satisfy the realizability and the second-order TVD property are applicable up to the third-order moment set. The second-order TVD conditions for the higher-order moments, fourth- and fifth-order moments, are conditionally satisfied, and they have shown the order of accuracy comparable to the lower order moments in the verification cases. 

\section*{Declaration of competing interest}
The authors declare that they have no known competing financial interests or personal relationships that could have appeared to influence the work reported in the present paper.

\section*{Acknowledgements}
The work was supported by the National Research Foundation of Korea (NRF) under the Grant Number NRF-2019K1A3A1A74107685 and NRF-2021R1A2C2092146.


\newpage
\bibliographystyle{elsarticle-num}
\biboptions{sort&compress}
\bibliography{variable_limiter}

\newpage

\newpage
\listoftables

\pagebreak
\clearpage	
\begin{table}
	\caption{Orders of accuracy for $m_0$ and $m_3$ obtained under the reugular initial number density function (NDF), the oscillating initial $\zeta$, and multi-modal initial NDF conditions using three different flux reconstruction schemes, respectively.}
	\centering
	\begin{tabular}{p{0.18\textwidth} p{0.1\textwidth} p{0.1\textwidth} p{0.1\textwidth} p{0.1\textwidth} p{0.1\textwidth} p{0.1\textwidth} } 
		\hline
		initial & \multicolumn{2}{l}{$\zeta$ simplified} & \multicolumn{2}{l}{equal limiter} & \multicolumn{2}{l}{variable limiter}  \\ 
		conditions
		& $m_0$ & $m_3$ & $m_0$ & $m_3$ & $m_0$ & $m_3$ \\ 
		\hline
		regular           & 1.93  & 1.93  & 0.98  & 1.27  & 1.92  & 1.90                        \\ 
		oscillating $\zeta$   & 1.92  & 1.92  & 1.65  & 1.35  & 1.92  & 1.93                       \\ 
		multi-modal       & 1.88  & 1.81  & 1.39  & 1.35  & 1.89  & 1.89                       \\ 
		\hline
	\end{tabular}
	\label{table_1}
\end{table}

\pagebreak
\clearpage
\begin{table}
	\caption{Orders of accuracy for $m_0$ and $m_3$ obtained under the 2D regular initial number density function (NDF) condition at $t=0.4$ and $t=0.8$ using three different flux reconstruction schemes, respectively.}
	\centering
	\begin{tabular}{p{0.18\textwidth} p{0.1\textwidth} p{0.1\textwidth} p{0.1\textwidth} p{0.1\textwidth} p{0.1\textwidth} p{0.1\textwidth} } 
		\hline
		time & \multicolumn{2}{l}{$\zeta$ simplified} & \multicolumn{2}{l}{equal limiter} & \multicolumn{2}{l}{variable limiter}  \\ 
		& $m_0$ & $m_3$ & $m_0$ & $m_3$ & $m_0$ & $m_3$ \\ 
		\hline
		$t = 0.4$   & 1.98  & 1.73  & 1.63  & 1.65  & 1.98  & 1.80                        \\ 
		$t = 0.8$   & 1.83  & 1.59  & 1.47  & 1.54  & 1.83  & 1.59                       \\ 
		\hline
	\end{tabular}
	\label{table_2}
\end{table}

\pagebreak
\clearpage

\newpage
\listoffigures

\pagebreak
\clearpage

\pagebreak
\clearpage	

\pagebreak
\clearpage
\begin{figure}
	\centering
	\includegraphics[width=\textwidth]{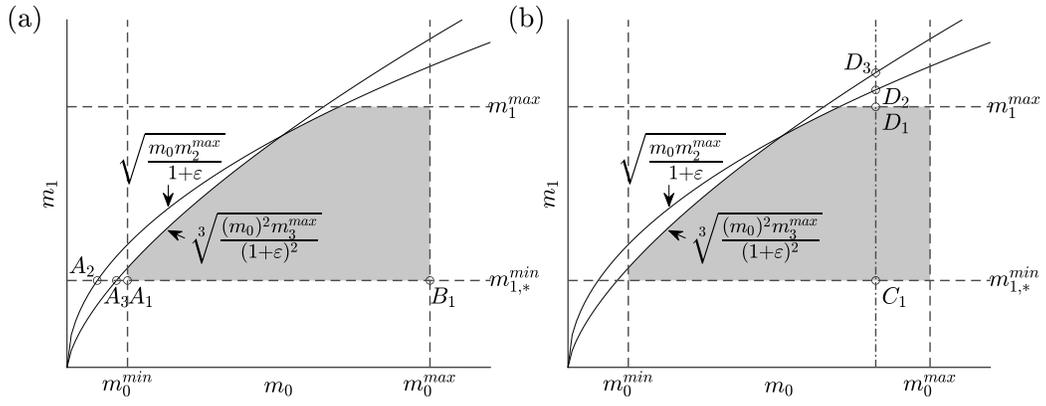}
	\caption{Schematic descriptions of the $m_0$ versus $m_1$ domain satisfying the realizability condition. The shaded regions represent the regions of the realizable $m_0$ and $m_1$. The four points $A_1, A_2, A_3,$ and $B_1$ in (a) limit the range of $m_0$ and the four points $C_1, D_1, D_2$, and $D_3$ in (b) limit the range of $m_1$ under the selected $m_0$.}
	\label{Fig_3}
\end{figure}

\pagebreak
\clearpage

\pagebreak
\clearpage

\pagebreak
\clearpage

\pagebreak
\clearpage

\pagebreak
\clearpage

\pagebreak
\clearpage

\pagebreak
\clearpage

\pagebreak
\clearpage

\pagebreak
\clearpage

\pagebreak
\clearpage

\pagebreak
\clearpage

\pagebreak
\clearpage

\pagebreak
\clearpage

\begin{algorithm} 	
	{\footnotesize			
		\caption{$\zeta$ slope reduction}	
		\label{alg_1}		
		\KwData{$\mathbf{m}_{N,i}^{n}$ and $\left(\mathbf{m}_{N,eo}^{n}\right)_{eo\in\{1,\ldots,N_{eo}\}}$}
		\KwResult{$\left(\mathbf{m}_{N,eo}^{mod}\right)_{eo\in\{1,\ldots,N_{eo}\}}$}
		$\left(\mathbf{m}_{N,eo}^{mod}\right)_{eo\in\{1,\ldots,N_{eo}\}}\leftarrow \left(\mathbf{m}_{N,eo}^{n}\right)_{eo\in\{1,\ldots,N_{eo}\}}$;\\
		$\mathbf{m}_{N}^{*}\leftarrow \max\left[\left(1+N_{eo}\right),\textmd{CFL}^{-1}\right]\mathbf{m}_{N,i}^{n}-\sum_{eo=1}^{N_{eo}}\mathbf{m}_{N,eo}^{mod}$;\\
		Compute $\mathcal{N}\left(\mathbf{m}_{N}^{*}\right)$ and $\mathcal{N}\left(\mathbf{m}_{N,i}^{n}\right)$;\\
		\If{$\mathcal{N}\left(\mathbf{m}_{N}^{*}\right)<\mathcal{N}\left(\mathbf{m}_{N,i}^{n}\right)$}{
			$\left(D_k\right)_{k\in\{0,\ldots,N\}} \leftarrow 1$;\\
			Compute $\left(\zeta_{0,i}^{n},\ldots,\zeta_{N,i}^{n}\right)$ and $\left(\zeta_{0,eo}^{n},\ldots,\zeta_{N,eo}^{n}\right)_{eo\in\{1,\ldots,N_{eo}\}}$;\\
			\For{$p \leftarrow 0$ \KwTo $\mathcal{N}\left(\mathbf{m}_{N,i}^{n}\right)$}{
				$\left(\zeta_{p,eo}^{mod}\right)_{eo\in\{1,\ldots,N_{eo}\}} \leftarrow \left(\zeta_{p,eo}^{n}\right)_{eo\in\{1,\ldots,N_{eo}\}}$;\\
				$\left(\zeta_{p+1,eo}^{mod},\ldots,\zeta_{N,eo}^{mod}\right)_{eo\in\{1,\ldots,N_{eo}\}}\leftarrow\left(\zeta_{p+1,i}^{n},\ldots,\zeta_{N,i}^{n}\right)$; \\
				Compute the moment sets $\left(\mathbf{m}_{N,eo}^{mod}\right)_{eo\in\{1,\ldots,N_{eo}\}}$;\\
				$\mathbf{m}_{N}^{*}\leftarrow \max\left[\left(1+N_{eo}\right),\textmd{CFL}^{-1}\right]\mathbf{m}_{N,i}^{n}-\sum_{eo=1}^{N_{eo}}\mathbf{m}_{N,eo}^{n}$;\\
				Compute $\mathcal{N}\left(\mathbf{m}_{N}^{*}\right)$;\\
				\If{$\mathcal{N}\left(\mathbf{m}_{N}^{*}\right)<\mathcal{N}\left(\mathbf{m}_{N,i}^{n}\right)$}{
					$D_p\leftarrow 0.5 D_p$;\\
					Compute $\left(\zeta_{p,eo}^{mod}\right)_{eo\in\{1,\ldots,N_{eo}\}}$ from Eq. \eqref{Eq_22};\\
					Compute the moment sets $\left(\mathbf{m}_{N,eo}^{mod}\right)_{eo\in\{1,\ldots,N_{eo}\}}$;\\
					$\mathbf{m}_{N}^{*}\leftarrow \max\left[\left(1+N_{eo}\right),\textmd{CFL}^{-1}\right]\mathbf{m}_{N,i}^{n}-\sum_{eo=1}^{N_{eo}}\mathbf{m}_{N,eo}^{mod}$;\\
					Compute $\mathcal{N}\left(\mathbf{m}_{N}^{*}\right)$;\\				
					\If{$\mathcal{N}\left(\mathbf{m}_{N}^{*}\right)<\mathcal{N}\left(\mathbf{m}_{N,i}^{n}\right)$}{
						$D_p\leftarrow 0$;\\
						$\left(\zeta_{p,eo}^{mod}\right)_{eo\in\{1,\ldots,N_{eo}\}}\leftarrow \zeta_{p,i}^{n}$;\\
				}}
			}
			Compute the moment sets $\left(\mathbf{m}_{N,eo}^{mod}\right)_{eo\in\{1,\ldots,N_{eo}\}}$;
		}
	}
\end{algorithm}

\pagebreak
\clearpage
\begin{algorithm}				
	{\footnotesize
		\caption{Variable flux limiter (part 1)}
		\label{alg_2}			
		\KwData{$\left(r_{k,e},m_{k,U},m_{k,D}\right)_{k\in\{0,\ldots,5\}}$}
		\KwResult{$\left(m_{k,e}\right)_{k\in\{0,\ldots,5\}}$}
		Compute $\left(m_{k,e}^{\min},m_{k,e}^{\max}\right)_{k\in\{0,\ldots,5\}}$ from Eq. \eqref{Eq_26} and Eq. \eqref{Eq_27};\\
		$\left(m_{k}^{\min},m_{k}^{\max}\right)_{k\in\{0,\ldots,5\}}\leftarrow\left(m_{k,e}^{\min},m_{k,e}^{\max}\right)_{k\in\{0,\ldots,5\}}$;\\
		$m_{1,*}^{\min} \leftarrow \max \left[m_{1}^{\min},\left(m_{2}^{\min}\right)^2\big/m_{3}^{\max}\right]$;\\ 
		$m_{0,*}^{\min} \leftarrow \max \left[\left(m_{1,*}^{\min}\right)^2\big/m_{2}^{\max},\sqrt[3]{\left(m_{1,*}^{\min}\right)^3\big/m_{3}^{\max}}\right]$;\\
		\uIf{$m_{0,*}^{\min}\le{m_{0}^{\max}}$ and $m_{1,*}^{\min}\le{m_{1}^{\max}}$}{	
			$m_0=
			\begin{cases}
				\max\left[m_0^{\min},m_{0,*}^{\min}\right] & \text{if}\quad{m_0^{mm}\le m_0^{sb}}\\
				m_0^{\max} & \text{if}\quad{m_0^{mm}> m_0^{sb}} \\
			\end{cases}
			$;\\
			$m_{0,\varepsilon}\leftarrow m_0$;\\
			$m_1=
			\begin{cases}
				m_{1,*}^{\min} & \text{if}\quad{m_1^{mm}\le m_1^{sb}}\\
				\min\left[m_1^{\max},\sqrt{m_0 m_2^{\max}},\sqrt[3]{(m_0)^2 m_3^{\max}}\right] & \text{if}\quad{m_1^{mm}> m_1^{sb}} \\
			\end{cases}
			$;\\	
			\If{$\sqrt{m_1 m_3^{\min}}>(m_1)^2/m_0$}{	
				$m_{0,\varepsilon}=
				\begin{cases}
					\max\left[m_0^{\min}/(1+\varepsilon),m_{0,*}^{\min}\right] & \text{if}\quad{m_0^{mm}\le m_0^{sb}}\\
					m_0^{\max}/(1+\varepsilon) & \text{if}\quad{m_0^{mm}> m_0^{sb}} \\
				\end{cases}
				$;\\
				$m_0\leftarrow (1+\varepsilon)m_{0,\varepsilon}$;\\	
				$m_1=
				\begin{cases}
					m_{1,*}^{\min} & \text{if}\quad{m_1^{mm}\le m_1^{sb}}\\
					\min\left[m_1^{\max},\sqrt{m_{0,\varepsilon} m_2^{\max}},\sqrt[3]{(m_{0,\varepsilon})^2 m_3^{\max}}\right] & \text{if}\quad{m_1^{mm}> m_1^{sb}} \\
				\end{cases}
				$;\\	
			}	
			$m_2=
			\begin{cases}
				\max\left[m_{2}^{\min},(m_1)^2/m_{0,\varepsilon}\right] & \text{if}\quad{m_2^{mm}\le m_2^{sb}}\\
				\min\left[m_{2}^{\max},\sqrt{m_1 m_3^{\max}}\right] & \text{if}\quad{m_2^{mm}> m_2^{sb}} \\
			\end{cases}
			$;\\
			$m_3=
			\begin{cases}
				\max\left[m_{3}^{\min},(m_2)^2/m_1\right] & \text{if}\quad{m_3^{mm}\le m_3^{sb}}\\
				m_{3}^{\max} & \text{if}\quad{m_3^{mm}> m_3^{sb}} \\
			\end{cases}
			$;\\	
	}}			
\end{algorithm}

\pagebreak
\clearpage
\setcounter{algocf}{1}
\begin{algorithm}			
	{\footnotesize
		\caption{Variable flux limiter (part 2)}
		\label{alg_2_2}			
		\Else{
			$\phi_{\min}\leftarrow\min\left[\phi_{mm}(r_0),\ldots,\phi_{mm}(r_3)\right]$;\\
			$\phi_{lower}(r)\leftarrow \max[0,\min(\phi_{\min},r)]$ in Eq. \eqref{Eq_25};\\
			Compute $\left(m_k\right)_{k\in\{0,\ldots,3\}}$ from Eq. \eqref{Eq_26} with $\gamma_k=0$;
		}
		Compute $\left(m_k\right)_{k\in\{4,5\}}$ from Eq. \eqref{Eq_26} with $\gamma_k=0$;\\
		Compute $\mathcal{N}\left(\mathbf{m}_{5}\right)$;\\
		\If{$\mathcal{N}\left(\mathbf{m}_{5}\right)\le5$}{
			$\left(m_k\right)_{k\in\{\mathcal{N}\left(\mathbf{m}_{5}\right),\ldots,5\}}\leftarrow \left(m_k^-\right)_{k\in\{0,\ldots,5\}}$ from Eq. \eqref{Eq_40}
		}	
		$\left(m_{k,e}\right)_{k\in\{0,\ldots,5\}}\leftarrow(m_k)_{k\in\{0,\ldots,5\}}$
	}
\end{algorithm}

\pagebreak
\clearpage

\begin{appendix}
	\section{Realizability condition for variable flux limiter}\label{Appendix_A}
	
	The feasible ranges of moments for the third-order moment set to simultaneously satisfy the second-order TVD and the realizability condition are given as follows:
	\begin{equation}
		\begin{gathered}
			m_0^{\min} \le m_0 \le m_0^{\max}, \\
			m_1^{\min} \le m_1 \le m_1^{\max}, \\
			\max\left[m_2^{\min},\frac{(m_1)^2}{m_0}\right] \le m_2 \le m_2^{\max},  \\
			\max\left[m_3^{\min},\frac{(m_2)^2}{m_1}\right] \le m_3 \le m_3^{\max}. \\
		\end{gathered}
		\label{Eq_A1}
	\end{equation}	
	Firstly, let us consider the range of $m_3$. The following condition should be satisfied for $m_3$ to exist in the proposed range:
	\begin{equation}
		\max\left[m_3^{\min},\frac{(m_2)^2}{m_1}\right]\le m_3^{\max} \quad \Longrightarrow \quad m_2 \le \sqrt{m_1 m_3^{\max}}.
		\label{Eq_A2}
	\end{equation}
	Accordingly, Eq. \eqref{Eq_A2} becomes the additional constraint in determining $m_2$ considering the existence of $m_3$. Therefore, the feasible range of $m_2$ is revised as follows:
	\begin{equation}
		\max\left[m_2^{\min},\frac{(m_1)^2}{m_0}\right] \le m_2 \le \min\left[m_2^{\max},\sqrt{m_1 m_3^{\max}}\right].
		\label{Eq_A3}
	\end{equation}
	Hence, the following condition should be satisfied for the revised range of $m_2$ to be acceptable:
	\begin{equation}
		\max\left[m_2^{\min},\frac{(m_1)^2}{m_0}\right] \le \min\left[m_2^{\max},\sqrt{m_1 m_3^{\max}}\right].
		\label{Eq_A4}
	\end{equation}
	
	This condition can be considered by dividing into two cases where $m_2^{\min}<(m_1)^2/m_0$ and $(m_1)^2/m_0\le m_2^{\min}$. If $m_2^{\min}<(m_1)^2/m_0$, the following constraint should be regarded in the range of $m_1$.
	\begin{equation}
		\begin{gathered}
			\frac{(m_1)^2}{m_0} \le \min\left[m_2^{\max},\sqrt{m_1 m_3^{\max}}\right] \quad \\ \Longrightarrow \quad 
			m_1 \le \min\left[\sqrt{m_0 m_2^{\max}},\sqrt[3]{(m_0)^{2} m_3^{\max}}\right].
		\end{gathered}
		\label{Eq_A5}
	\end{equation}
	In addition, the assumption of $m_2^{\min}<(m_1)^2/m_0$ implies the necessity of $\sqrt{m_0 m_2^{\min}} < m_1$. Therefore, the constraint in Eq. \eqref{Eq_A5} can be rewritten as follows: 
	\begin{equation}
		\sqrt{m_0 m_2^{\min}} < m_1 \le \min\left[\sqrt{m_0 m_2^{\max}},\sqrt[3]{(m_0)^{2} m_3^{\max}}\right].
		\label{Eq_A6}
	\end{equation}
	In Eq. \eqref{Eq_A6}, at the point where it satisfies the condition of $\sqrt{m_0 m_2^{\min}}=\sqrt[3]{(m_0)^{2} m_3^{\max}}$, $m_0$ becomes $\left(m_2^{\min}\right)^{3}\big/\left(m_3^{\max}\right)^{2}$, and the corresponding $m_1$ becomes $\left(m_2^{\min}\right)^{2}\big/m_3^{\max}$. The constraint of $\left(m_2^{\min}\right)^{2}\big/m_3^{\max}<m_1$ is implied on the assumption since the condition of $\sqrt{m_0 m_2^{\min}}<\sqrt[3]{(m_0)^{2} m_3^{\max}}$ is only valid in the range where $\left(m_2^{\min}\right)^{3}\big/\left(m_3^{\max}\right)^{2}<m_0$. Consequently, the feasible range of $m_1$ where $m_2$ and $m_3$ can be determined is provided as follows with the assumption of $m_2^{\min}<(m_1)^2/m_0$:
	\begin{equation}
		\frac{\left(m_2^{\min}\right)^{2}}{m_3^{\max}} < m_1 \le \min\left[\sqrt{m_0 m_2^{\max}},\sqrt[3]{(m_0)^{2} m_3^{\max}}\right].
		\label{Eq_A7}
	\end{equation}
	If $(m_1)^2/m_0\le m_2^{\min}$, the following constraint should be regarded in the range of $m_1$:
	\begin{equation}
		m_2^{\min} \le \min\left[m_2^{\max},\sqrt{m_1 m_3^{\max}}\right] \quad \Longrightarrow \quad 
		\frac{\left(m_2^{\min}\right)^{2}}{m_3^{\max}} \le m_1.
		\label{Eq_A8}
	\end{equation}
	In addition, the assumption of $(m_1)^2/m_0\le m_2^{\min}$ implies the following constraint:
	\begin{equation}
		\begin{gathered}
			\frac{(m_1)^2}{m_0} \le m_2^{\min} \le \min\left[m_2^{\max},\sqrt{m_1 m_3^{\max}}\right] \quad \\ \Longrightarrow \quad 
			m_1 \le \min\left[\sqrt{m_0 m_2^{\max}},\sqrt[3]{(m_0)^{2} m_3^{\max}}\right].
		\end{gathered}
		\label{Eq_A9}
	\end{equation}
	
	Consequently, the feasible range of $m_1$ where $m_2$ and $m_3$ can be determined is provided below based on the assumption of $(m_1)^2/m_0\le m_2^{\min}$:
	\begin{equation}
		\frac{\left(m_2^{\min}\right)^{2}}{m_3^{\max}} \le m_1 \le \min\left[\sqrt{m_0 m_2^{\max}},\sqrt[3]{(m_0)^{2} m_3^{\max}}\right].
		\label{Eq_A10}
	\end{equation}
	The feasible ranges of $m_1$ for two different assumptions are identical except for the equality case, $(m_1)^2/m_0 = m_2^{\min}$. Therefore, the feasible range of $m_1$ in Eq. \eqref{Eq_A1} is revised as follows:
	\begin{equation}
		\max\left[m_1^{\min},\frac{\left(m_2^{\min}\right)^2}{m_3^{\max}}\right] \le m_1 
		\le \min\left[m_1^{\max},\sqrt{m_0 m_2^{\max}},\sqrt[3]{\left(m_0\right)^2 m_3^{\max}}\right].
		\label{Eq_A11}
	\end{equation}
	Here, we consider the case where $\Delta_2$ becomes 0 (i.e., $m_2=(m_1)^2/m_0$). This case occurs when the moment set belongs to the boundary of the moment space. Thus,  $m_3$ is forced to make $\Delta_3=0$ (i.e., $m_3=(m_2)^2/m_1$). Therefore, the additional constraint of $m_3^{\min}\le(m_2)^2/m_1$ is involved for $m_3$ to exist within the proposed range in Eq. \eqref{Eq_30}. Under the condition of $\Delta_2=0$, the additional constraint is rewritten as follows: 
	\begin{equation}
		\sqrt{m_1 m_3^{\min}}\le m_2 \quad \Longrightarrow \quad \sqrt{m_1 m_3^{\min}}\le\frac{(m_1)^2}{m_0}.
		\label{Eq_A12}
	\end{equation}
	If the condition of Eq. \eqref{Eq_A12} is satisfied, $\Delta_2=0$ and $\Delta_3=0$ are allowed (i.e., $m_2=(m_1)^2/m_0$ and $m_3=(m_2)^2/m_1$). However, if not, to prevent the problem of belonging to the boundary, a small value $\varepsilon$ is introduced in Eq. \eqref{Eq_A2} as follows:
	\begin{equation}
		\max\left[m_2^{\min},(1+\varepsilon)\frac{(m_1)^2}{m_0}\right] \le m_2 \le \min\left[m_2^{\max},\sqrt{m_1 m_3^{\max}}\right].
		\label{Eq_A13}
	\end{equation}
	
	As a result, the issue of belonging to the boundary that comes from the condition of $m_2=(m_1)^2/m_0$ is prevented. In the present study, the value of $\varepsilon$ is set to $10^{-6}$. The range of $m_1$ in Eq. \eqref{Eq_A11} is revised by considering $\varepsilon$ as follows:
	\begin{equation}
		\begin{gathered}
			m_{1,*}^{\min} \le m_1 
			\le \min\left[m_1^{\max},\sqrt{\frac{m_0 m_2^{\max}}{1+\varepsilon}},\sqrt[3]{\frac{\left(m_0\right)^2 m_3^{\max}}{(1+\varepsilon)^2}}\right], \\ 
			\textmd{where} \quad m_{1,*}^{\min}=\max\left[m_1^{\min},\frac{\left(m_2^{\min}\right)^2}{m_3^{\max}}\right].
		\end{gathered}
		\label{Eq_A14}
	\end{equation}
	
	Fig. \ref{Fig_3} shows  the constraint of Eq. \eqref{Eq_A14} graphically with $m_0$ on the interval $\left[m_0^{\min},m_0^{\max}\right]$. The feasible region of the pair $(m_0,m_1)$ is shaded in Fig. \ref{Fig_3}. Therefore, the available range of $m_0$ is determined by the four points denoted as $A_1$, $A_2$, $A_3$, and $B_1$ in the plot. These points are the intersection points of the functions shown in Fig. \ref{Fig_3}(a), and they have the following values:
	\begin{equation}
		\begin{gathered}
			A_{1}=\left(m_{0}^{\min}, m_{1,*}^{\min}\right),\quad
			A_{2}=\left((1+\varepsilon)\frac{\left(m_{1,*}^{\min}\right)^{2}}{m_{2}^{\max}}, m_{1,*}^{\min}\right),\\
			A_{3}=\left((1+\varepsilon)\sqrt{\frac{\left(m_{1,*}^{\min}\right)^{3}}{m_{3}^{\max}}}, m_{1,*}^{\min}\right),\quad
			B_{1}=\left(m_{0}^{\max}, m_{1,*}^{\min}\right).
		\end{gathered}
		\label{Eq_A15}
	\end{equation}
	Accordingly, the feasible range of $m_0$ is provided as follows:
	\begin{equation}
		\max \left[m_{0}^{\min}, (1+\varepsilon)\frac{\left(m_{1,*}^{\min}\right)^{2}}{m_{2}^{\max}}, 
		(1+\varepsilon)\sqrt{\frac{\left(m_{1,*}^{\min}\right)^{3}}{m_{3}^{\max}}}\right] \le m_{0} \le m_{0}^{\max},
		\label{Eq_A16}
	\end{equation}
	where $m_{1,*}^{\min}=\max\left[m_1^{\min},\left(m_2^{\min}\right)^2/m_3^{\max}\right]$. Subsequently, if $m_0$ is determined in the range of Eq. \eqref{Eq_A16}, the value of $m_1$ can be determined along the dot-dashed line in Fig. \ref{Fig_3}(b). Therefore, the available range of $m_0$ is determined by the four intersection points denoted as $C_1$, $D_1$, $D_2$, and $D_3$ in the plot, and they have the following values:
	\begin{equation}
		\begin{gathered}
			C_{1}=\left(m_{0}, m_{1,*}^{\min}\right),\quad
			D_{1}=\left(m_{0}, m_{1}^{\max}\right),\\
			D_{2}=\left(m_{0}, \sqrt{\frac{m_0 m_2^{\max}}{1+\varepsilon}}\right),\quad
			D_{3}=\left(m_{0}, \sqrt[3]{\frac{\left(m_0\right)^2 m_3^{\max}}{(1+\varepsilon)^2}}\right).
		\end{gathered}
		\label{Eq_A17}
	\end{equation}
	
	This graphical approach leads to the identical range of Eq. \eqref{Eq_A14}. Consequently, the feasible ranges of the moments are presented below to satisfy the second-order TVD and the realizability condition concurrently.
	\begin{equation}
		\begin{gathered}
			\max \left[{m_0}^{\min}, (1+\varepsilon)\frac{\left(m_{1,*}^{\min}\right)^{2}}{m_{2}^{\max}}, 
			(1+\varepsilon)\sqrt{\frac{\left(m_{1,*}^{\min}\right)^{3}}{m_{3}^{\max}}}\right] \le m_{0} \le m_{0}^{\max}, \\
			m_{1,*}^{\min} \le m_1 
			\le \min\left[m_1^{\max},\sqrt{\frac{m_0 m_2^{\max}}{1+\varepsilon}},\sqrt[3]{\frac{\left(m_0\right)^2 m_3^{\max}}{(1+\varepsilon)^2}}\right], \\
			\max\left[m_2^{\min},(1+\varepsilon)\frac{(m_1)^2}{m_0}\right] \le m_2 \le \min\left[m_2^{\max},\sqrt{m_1 m_3^{\max}}\right], \\
			\max\left[m_3^{\min},\frac{(m_2)^2}{m_1}\right] \le m_3 \le m_3^{\max}, \\
		\end{gathered}
		\label{Eq_A18}
	\end{equation}
	where $m_{1,*}^{\min}=\max\left[m_1^{\min},\left(m_2^{\min}\right)^2/m_3^{\max}\right]$.

\end{appendix}

\pagebreak
\clearpage

\thispagestyle{empty}
\section*{Highlights}
\begin{sloppypar}
\begin{itemize}

    \item{A novel realizable 2nd-order advection method is developed for moments transport.}
		
    \item{A variable flux limiter is devised to satisfy the 2nd-order TVD property.}

	\item{Conditions to satisfy realizable 2nd-order TVD are proposed for 3rd-order moment set.}
	
\end{itemize}
\end{sloppypar}
\pagebreak
\clearpage

\end{document}